\documentclass[aps,pra,twoside,twocolumn,10pt]{revtex4-1}
\usepackage{epsfig,newlfont,amssymb,amsfonts,amsmath,bm,subfigure,palatino,mathtools,amsthm,braket,soul,enumitem,color,graphics,graphicx,times,physics,comment}
\usepackage{float}
\usepackage[normalem]{ulem}

\usepackage[colorlinks=true, citecolor=blue, urlcolor=blue ]{hyperref}

\begin{document}
\title{Quantum resources in Harrow-Hassidim-Lloyd algorithm }

\author{Pradeep Kumar$^{a,b}$, Tanoy Kanti Konar$^{a}$, Leela Ganesh Chandra Lakkaraju$^{a}$, Aditi Sen(De)$^{a}$}
\affiliation{$^{a}$Harish-Chandra Research Institute, A CI of Homi Bhabha National Institute,  Chhatnag Road, Jhunsi, Allahabad - 211019, India\\$^{b}$ Institute for Quantum Science and Technology, University of Calgary, Alberta T2N 1N4, Canada}

\begin{abstract}

Quantum algorithms have the ability to reduce runtime for executing tasks beyond the capabilities of classical algorithms. Therefore, identifying the resources responsible for quantum advantages is  an interesting endeavour. We prove that nonvanishing quantum correlations, both bipartite and genuine multipartite entanglement, are required for  solving nontrivial linear systems of equations in the Harrow-Hassidim-Lloyd (HHL) algorithm. Moreover, we find a nonvanishing \(l_1\)-norm quantum coherence of the entire system and the register qubit which turns out to be related to the success probability of the algorithm. Quantitative analysis of the quantum resources reveals that while a significant amount of bipartite entanglement is generated in each step and required for this algorithm, multipartite entanglement content is inversely proportional to the  performance indicator. In addition, we report that when imperfections chosen from Gaussian distribution are incorporated in controlled rotations, multipartite entanglement increases with the strength of the disorder, albeit error also increases while bipartite entanglement and coherence decreases,  confirming the beneficial role of bipartite entanglement and coherence in this algorithm.

\end{abstract}

\maketitle

\section{Introduction}
\label{sec:introduction}

Quantum computers  have the potential to revolutionize data processing and other computational tasks in terms of time complexity and memory synchronization \cite{nielsen_chuang_2010}. It has been demonstrated that the speed up in computation and benefits obtained in many quantum protocols, including quantum communication \cite{quantum_crypto_gisin_rmp_2002, aditi_quantum_comm_review_2011} and quantum sensing \cite{quantum_sensing_review_2017}, are dependent on nonclassical properties of composite or single quantum systems, such as entanglement \cite{horodecki2009},  coherence \cite{streltsov_rmp_2017} etc \cite{kendon_qic_2006, coherence_algo_2019} . For example, quantum algorithms such as  Grover's search \cite{grover_arxiv_1996, grover_prl_1997, grover_entanglement_PRA_2013, pan_qic_2017}, Deutsch-Jozsa \cite{Deutsch_Jozsa,cleve_1998,gangopadhyay_qip_2018}, Shor's factorization  \cite{shor_1994,lu_prl_2007,azuma_jmo_2018}, Bernstein-Vazirani \cite{entanglement_bernstein_pra_2022} algorithms exploit quantum correlations at various stages to improve their performance. Moreover, multipartite entanglement also turns out to be crucial ingredient in quantum error correction \cite{shor_pra_1995}, quantum algorithms \cite{bruss_pra_2011} and measurement-based quantum computation \cite{raussendorf_prl_2001}. In addition to theoretical advancements, these algorithms have also been demonstrated in laboratories using nuclear magnetic resonance (NMR) techniques \cite{ermakov_pra_2002, Xin_cps_2018, xin_pra_2020,xin_prl_2021}, trapped ions \cite{Gulde_nature_2003}, neutral atoms \cite{neutralatom_phaseestimation_2022}, superconducting processors \cite{zheng_prl_2017,takita_prl_2021}, photonics \cite{shor_experi_photonics_2007}, silicon qubits \cite{silicon_experimental_deutsch_grover_2018}, photonic quantum logic gates \cite{lanyon_prl_2007} and also using supervised machine learning \cite{Havlicek_natute_2019}.

Nonlocal  quantum correlations have already been established as the key resource  in several communication tasks like quantum teleportation,  dense coding and quantum key distribution \cite{bennett_prl_1992, benett_prl_1993, ekert1991} which cannot be accomplished using their classical counterparts. Similarly, in recent years,  entanglement has been shown to be a detector of physical phenomena including quantum phase transition \cite{amico_fazio_rmp_2008, aditi_advanves_physics_2007}, localization-delocalization transitions \cite{bloch_rmp_2019}.  However,
the relevance of nonclassicality in the speed-up of quantum computation has yet to be determined.
 Recently, quantum coherence is proven to be vital in obtaining quantum advantages in the Bernstein-Vazirani algorithm \cite{entanglement_bernstein_pra_2022} as well as in the Grover's \cite{ent_cohe_grover_2019}, Deutsch-Jozsa algorithm \cite{hillery_pra_2016}. Similarly, nonvanishing  bipartite and multipartite entanglement in Deutsch-Josza  \cite{silicon_experimental_deutsch_grover_2018}, Grover's search \cite{grover_entanglement_PRA_2013, ent_cohe_grover_2019, global_ent_grover_2017, multi_ent_grover_2012}, Simon \cite{bruss_pra_2011} and Shor's \cite{shors_entanglement_2006} algorithms have also been reported. 



In this paper,  we focus on the Harrow-Hassidim-Lloyd (HHL) algorithm \cite{HHL_PRL_2009} which demonstrates exponential speed up in solving linear systems of equations,  including partial differential equations like  Poisson's equation \cite{cao_njp_2013}, when compared to classically known algorithms. Linear systems of equations are commonly used to represent and analyse a wide range of phenomena, including fluid dynamics, electrical circuits, quantum physics, and optimisation problems.  Hence, finding solution of these equations in an efficient way can be crucial for  making accurate predictions in various scientific and engineering disciplines. The HHL algorithm, a quantum version of solving these equations, performs the task by encoding the input vector and the matrix representing the linear system into a quantum state, which is then manipulated using quantum gates to obtain a new state containing the solution vector \cite{HHL_PRL_2009, Yudong_mp_2012, morrell_arxiv_2023}.  Furthermore,  quantum coherence and entanglement have also been computed by considering specific linear systems of equations \cite{hwang_arxiv_2022,ent_cohe_HHL_2022} to understand their role in the algorithm. 
 It has been demonstrated experimentally in a four-qubit NMR \cite{pan_pra_2014}, photonics \cite{pan_prl_photonics_2013, aspuru_photonics2_2014}, and superconducting processors  \cite{Supercond_hhl_2020, zheng_prl_2017} by solving two-dimensional systems of linear equations. 

In this work, we address the role of quantum resources required to execute different steps of the HHL algorithm which deals with an arbitrary dimensional systems of linear equations.  To capture the nonclassicality,  we divide the entire state into three parts -- (i) a quantum register of \(n\) qubits referred to as \(\mathbf{\Lambda}\), (ii) a register of \(n_b\) qubits marked as \(\mathbf{U}\) and (iii) a single auxiliary qubit. Specifically, we analytically prove that  genuine multipartite entanglement (GME) quantified via  generalized geometric measure (GGM) \cite{sen_pra_2010} of this tripartite state has to be nonvanishing after the control rotation performed in the algorithm while the closed form expressions for bipartite entanglement between different subsystems, \(\mathbf{\Lambda}\) and \(\mathbf{U}\) or \(\mathbf{U}\) and \(\mathbf{R}\) is always nonvanishing for successful implementation. Interestingly, by considering two- and three-dimensional systems of linear equations,  we demonstrate that the time complexity, a performance indicator, of the HHL algorithm increases with the decrease of bipartite entanglement, produced after the controlled rotation while the opposite picture emerges for GME, thereby establishing the detrimental effect of GME in the algorithm. 
On the other hand, quantum coherence \cite{streltsov_rmp_2017} for the entire systems and the register qubit is directly related to the performance indicator and the latter is also connected to the success probability of the protocol.


To examine the robustness of the algorithm, we introduce errors  into the control rotation  by deviating the angle chosen randomly  from the Gaussian distribution (see similar works in which  the effects of  noise or disorder on  quantum algorithms  have been studied \cite{maiti_arxiv_2022,gupta_arxiv_2023}).  We observe that the quenched average  error along with the GGM of the state after control rotation gets increased with the increase in the strength of the disorder. However, quenched average bipartite entanglement diminishes with the increase of disorder which is consistent with the errors in solutions.
It establishes that, while two- and multiparty entanglements are required for executing the HHL algorithm, a significant degree of GME can have a negative impact on the algorithm's success. However, 
  successful execution necessitates a substantial amount of bipartite entanglement and coherence.


The structure of the paper is as follows: in the Sec.  \ref{sec:hhl_algorithm}, we provide the generic stages of the HHL algorithm and the corresponding states to be prepared. In Sec. \ref{sec:roleentvscoh}, we prove the presence of nonvanishing quantum resources in the algorithm while the connection between quantum features and the performance is established in Sec. \ref{sec:connect}. Sec. \ref{sec:disorder} deals the effects of imperfections on quantum resources and quantum algorithms. The conclusion is presented in Sec. \ref{sec:conclu}.


\section{Harrow-Hassidim-Lloyd algorithm: Step by step }
\label{sec:hhl_algorithm}

The HHL algorithm implemented on a quantum system solves a set of linear equations, written in a matrix form as $\mathbf{A}\vec{x} = \vec{b}$. Here $\vec{x}$ is the vector which contains variables whose values we intend to find, \(\mathbf{A}\) is a $N\cross N$  Hermitian matrix containing coefficients of the equations and \(\vec{b}\) is the vector containing constants. Notice that the quantum version can provide an exponential speed-up in finding the solutions in comparison with the existing algorithms available for a classical computer \cite{HHL_PRL_2009}.

In order to obtain the solution of the linear equations, we scale $\vec{b}$ to unit length such that it can be implemented as a valid quantum state in the quantum circuit. Moreover, $\mathbf{A}$ is a Hermitian matrix although for a non-Hermitian matrix $(A^\prime)$, a corresponding Hermitian matrix can be calculated as $\mathbf{A} = \begin{bmatrix}
    0 & A^\prime \\ (A^\prime)^\dagger & 0 
\end{bmatrix}$ which can then be used in the protocol. The solution of the equations realized as \(\vec{x}\),  written in a quantum domain reads as 
$
    \ket{x} = \mathbf{A^{-1}} \ket{b}.
$ 
Thus the algorithm involves embedding the inverse eigenvalues of $\mathbf{A}$ into a quantum state such that the state $|x\rangle$ can be found at the end of the protocol, which is illustrated as follows.  


\begin{figure}
\includegraphics[width=1\linewidth]{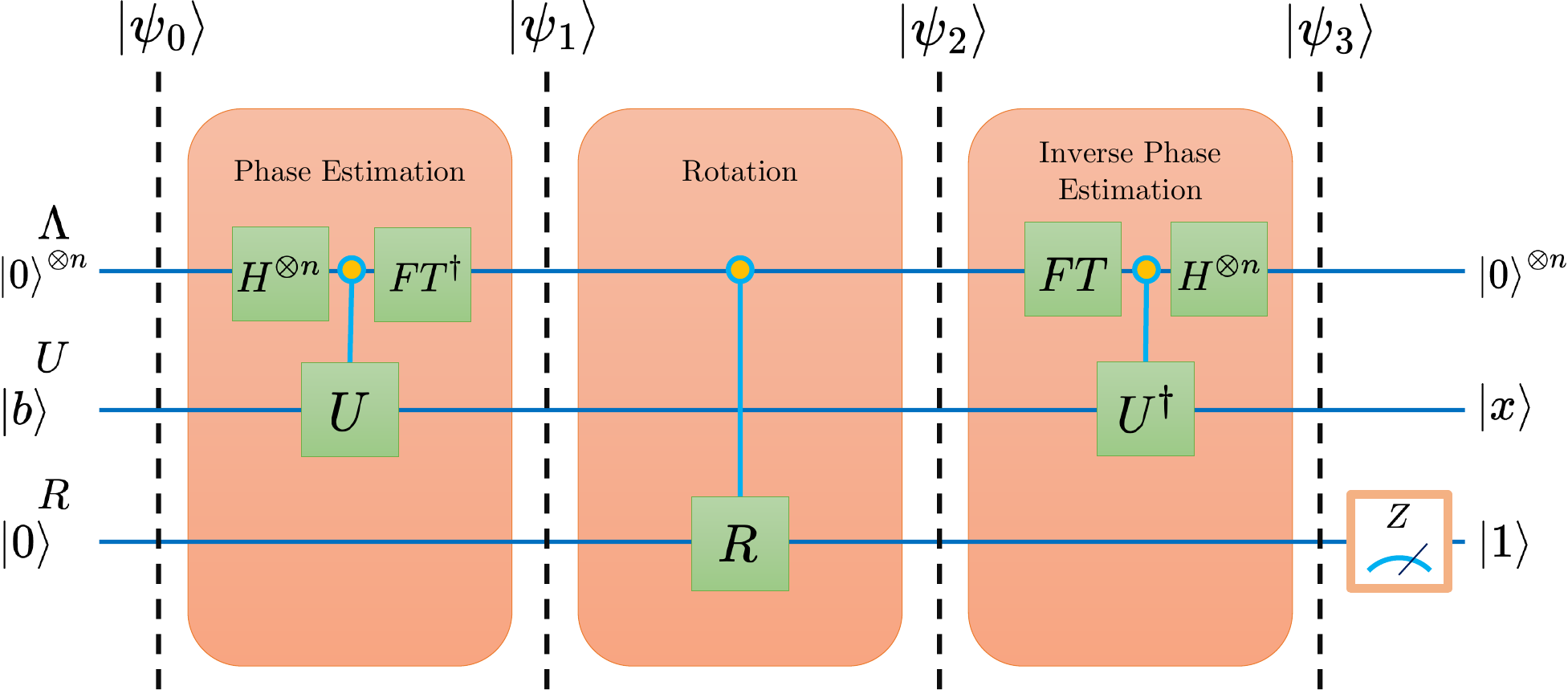}
\caption{\label{fig:schematic} Schematic diagram of the HHL Algorithm. The entire system can be divided into three subsystems, a quantum register of \(n\) qubits (\(\mathbf{\Lambda}\)),  a register of \(n_b\) qubits (\(\mathbf{U}\)) and  a single auxiliary qubit (\(\mathbf{R}\)). In the first step, we apply the phase estimation algorithm, which consists of initially applying a Hadamard gate on the $\Lambda$ subsystem, followed by a unitary operation ($e^{i \mathbf{A} t}$), and then an inverse Fourier transform. In the second step,  a controlled rotation that rotates the auxiliary qubit based on the eigenvalues of the linear system is executed which leads to the state, \(|\psi_2\rangle\). In the final step, we perform an inverse phase transform, resulting to a state \(|\psi_3\rangle\). Finally, measuring the auxiliary subsystem, \(\mathbf{R}\) in the computational basis, and upon post-selecting $\ket{1}$ as the measurement result, we obtain the solution. }
\end{figure}

{\it Step 0. - Initialization of the quantum registers.} Let us prepare a state consisting of three subsystems, a register of \(n\) qubits of working memory $\ket{0}^{\otimes n}$, denoted as  $\mathbf{\Lambda}$, a register of $n_b$ qubits, representing the subsystem $\mathbf{U}$, and a single auxiliary qubit initialised in $\ket{0}$, referred to as $\mathbf{R}$. Thus, the entire state is represented as 
\begin{equation}
    \ket{\psi_{in}} = \ket{0}^{\otimes n}_{\mathbf{\Lambda}} \otimes \ket{0}^{\otimes n_b}_{\mathbf{U}} \otimes \ket{0}_{\mathbf{R}},  
\end{equation} 
which is written in the partition $ \mathbf{\Lambda} : \mathbf{U} : \mathbf{R}$. Since all the operations in the algorithm are performed in this partition,   quantum correlations or coherence computed in this partition can possibly contribute  to achieve quantum advantage. 
The subsystem $\ket{0}^{\otimes n_b}$  encodes  the coefficients of $\vec{b}$, thereby producing the state as
\begin{equation}
    \ket{\psi_{0}} = \ket{0}^{\otimes n}_{\mathbf{\Lambda}} \otimes \ket{b}_{\mathbf{U}} \otimes \ket{0}_{\mathbf{R}} = \ket{0}^{\otimes n}_{\mathbf{\Lambda}} \otimes \sum_{i = 0}^N \beta_i |u_i\rangle \otimes \ket{0}_{\mathbf{R}},
    \label{eq:0step}
\end{equation} 
where \(N\) is the linear system of equations. Note that the constant vector is decomposed in the eigenvector basis of $\mathbf{A}$ to embed the corresponding eigenvalues into the $\mathbf{\Lambda}$ register.

{\it Step 1. Encoding of the eigenvalues.} Let us encode eigenvalues $\lambda_i$ of  \(\mathbf{A}\) in the subsystem $\mathbf{\Lambda}$ by utilizing quantum phase estimation (QPE) \cite{nielsen_chuang_2010}. In order to do that, we create a state which contains relative phases as the eigenvalues. The unitary operator involved in this protocol can be performed by Hamiltonian simulation, i.e., by letting the system evolve for a particular time $t$ \cite{Berry_cmp_2007, Barry_dynamics_2, Barry_dynamics_3} as $V = e^{i\textbf{A}t}$, which is then applied on $\ket{b} = \sum_{i=1}^{N} \beta_{i} \ket{u_{i}}$
$
V \sum_{i=1}^{N} \beta_{i} \ket{u_{i}} = \sum_{i=1}^{N} \beta_{i} e^{i\textbf{A}t}\ket{u_{i}} = \sum_{i=1}^{N} \beta_{i} e^{i\lambda_{i}t}\ket{u_{i}}.
$

The quantum phase estimation  has three components, namely the application of Hadamard gates, controlled rotation, and inverse quantum Fourier transform. The QPE operation thus encodes the phase value in the empty qubit, written mathematically as 
\begin{equation}
QPE[\sum_{i=1}^{N}\ket{0}^{\otimes n}_{\mathbf{\Lambda}} e^{2 \pi i \phi} \ket{u_i}_{\mathbf{U}}] = \sum_{i=1}^{N}\ket{2^{n}\phi}_{\mathbf{\Lambda}} \beta_i \ket{u_i}_{\mathbf{U}}. 
\end{equation}
The eigenvalues of $\mathbf{A}$ are created as the relative phases of the state in the subsystem, $\mathbf{U}$ and the QPE can be performed to rotate the register containing $|0\rangle$ in the subsystem $\mathbf{\Lambda}$ to reflect the eigenvalues as states. Finally, the Hadamard gates are applied to the subsystem $\mathbf{\Lambda}$, and the controlled rotation gates are applied on both $\mathbf{\Lambda}$ and  $\textbf{U}$ such that the rotation from $|0\rangle$ is heralded on the qubit in $\mathbf{U}$.  

The QPE estimates the phase $\phi$ from the relation $2\pi i \phi =  i\lambda_{i}t$ as $\phi = \lambda_{i}t/2\pi$ which is then encoded in the n-qubit register as $\ket{2^{n}\frac{\lambda_{i}t}{2\pi}}_{\mathbf{\Lambda}}$. 
Note that $\lambda_i$s are usually not integers and hence, we choose \(t\) in such a way that $\tilde{\lambda}_{i} = 2^{n} \lambda_{i} t / 2\pi$ are integers, which are scaled version of $\lambda_{i}$. If the eigenvalues are irrational numbers, the precision is decided by the number of qubits(\(n\)) in the register. For ease of calculation and presentation, we will only consider integer eigenvalues. After the QPE, the state of the system reduces to
\begin{equation}
\ket{\psi_{1}} = \sum_{i=1}^{N}\ket{\lambda_{i}}_{\mathbf{\Lambda}} \otimes \beta_{i} \ket{u_{i}}_{\mathbf{U}} \otimes \ket{0}_{\mathbf{R}},
\label{eq:1ststep}
\end{equation}

{\it Step 2. Encoding in the auxiliary qubit.} The information of \(\mathbf{A}^{-1}\) is encoded in the auxiliary qubit by applying controlled rotation $R(\lambda^{-1})$ on that qubit. The resulting state becomes
\begin{equation} 
\ket{\psi_{2}} = \sum_{i=1}^{N}\ket{\lambda_{i}}_{\mathbf{\Lambda}} \otimes \beta_{i} \ket{u_{i}}_{\mathbf{U}} \otimes \Big (\sqrt{1-\frac{C^{2}}{\lambda_{i}^{2}}}\ket{0} + \frac{C}{\lambda_{i}}\ket{1}\Big )_{\mathbf{R}},
\label{eq:step2}
\end{equation}
which is multipartite entangled as we will prove explicitly, thereby confirming its importance in the algorithm. The transformation of $\ket{0}_{\mathbf{R}}$ to $\Big (\sqrt{1-\frac{C^{2}}{\lambda_{i}^{2}}}\ket{0} + \frac{C}{\lambda_{i}}\ket{1}\Big )_{\mathbf{R}}$ occurs  due to the rotation, represented as
\begin{equation}
    R(\lambda^{-1}) = R_{y}(\theta) = \begin{bmatrix}
  \cos\frac{\theta}{2} & -\sin\frac{\theta}{2} \\[1ex]
  \sin\frac{\theta}{2} & \cos\frac{\theta}{2} \\
\end{bmatrix},
\end{equation}
\noindent
where $\theta = 2\sin^{-1}(\frac{C}{\lambda})$. Here \(C\) is a circuit constant which can take maximum value up to \(1\) where the lowest eigenvalue of the matrix \(\mathbf{A}\) is scaled to \(1\). Note that we consider a constant which is of the order of $\mathcal{O}(\frac{1}{\kappa})$, with $\kappa$ being the ratio between the highest and lowest eigenvalues of $\mathbf{A}$.

{\it Step 3. Reverse QPE and read out.} Applying reverse quantum phase estimation,  the corresponding state now becomes
\begin{equation} 
\ket{\psi_{3}} = \ket{0}^{\otimes n}_{\mathbf{\Lambda}} \otimes \sum_{i=1}^{N}  \beta_{i} \ket{u_{i}}_{\mathbf{U}} \otimes \Big (\sqrt{1-\frac{C^{2}}{\lambda_{i}^{2}}}\ket{0} + \frac{C}{\lambda_{i}}\ket{1}\Big )_{\mathbf{R}},
\label{eq:step3}
\end{equation}
which can then be used to read out the results. For that purpose, measurement is performed on the auxiliary qubit in the computational basis and the post-selection of an outcome $\ket{1}$ results to the correct result  as
\begin{equation}
\ket{\psi_{\vec{x}}} = {\frac{1}{\sqrt{\mathcal{SP}}}} \sum_{i=n}^{N} C\frac{\beta_{i}}{\lambda_{i}}\ket{u_{i}}.
\label{eq:soln}
\end{equation}
{
Here the success probability can be calculated as $ \mathcal{SP} = \sum_{i=1}^{N}\beta_i^2\frac{C^2}{\lambda_i^2}$. In other words, the probability of obtaining the state, $\ket{\psi_{\vec{x}}}$ after the measurement is given by $\mathcal{SP}$, which is a figure of merit  utilized in the subsequent sections to assess the performance of the algorithm. It is important to note that accessing the individual solutions, i.e., the elements of the state $\ket{\psi_{\vec{x}}}$ can be tomographically hard, although performing any arithmetic using the vector can be computationally feasible  \cite{babukhin_pra_2023}, and hence post selection is not required in order to achieve the expectation value of any observable.}

\section{Entanglement and coherence in  HHL Algorithm}
\label{sec:roleentvscoh}

It is always intriguing to identify quantum features which lead to advantages in quantum algorithms. For example, entanglement and quantum coherence are found to play important roles in different stages of algorithms including Grover's search \cite{grover_entanglement_PRA_2013}, Shor's factorization  \cite{shors_entanglement_2006},  and Bernstein-Vazirani algorithms \cite{entanglement_bernstein_pra_2022}. We will now establish that multipartite as well as bipartite entanglement and coherence created in the HHL algorithm are crucial.  We will accomplish this in two steps - first, we prove analytically that the algorithm can provide nontrivial solution when quantumness in the form of entanglement and coherence is nonvanishing. Secondly, we quantitatively assess the entanglement and coherence content present in the system and its connection with complexity or success probability of the HHL algorithm. 

The resources studied  here are genuine multipartite entanglement measure, called generalized geometric measure, \(\mathcal{E}\) \cite{sen_pra_2010, ggm_Barnum2001Aug, ggm_Goldbart2003, ggm_Shimony1995Apr}, bipartite entanglement measure, namely logarithmic negativity ($\mathcal{LN}$) \cite{vidal_logneg_2002, plenio2005}, and $l_1$-norm quantum coherence measures \cite{baumgratz_prl_2014,streltsov_rmp_2017}. 
Let us define briefly multipartite and bipartite entanglement measures used in this article. \\

{\it GGM.} The generalized geometric measure of an N-party pure quantum state, $ \ket{\psi_{N}}$, is a computable entanglement measure that can quantify genuine multiparty entanglement content of the state, i.e., it vanishes for all non-genuinely multipartite pure states which is product across one of the bipartition \cite{sen_pra_2010, ggm_Barnum2001Aug, ggm_Goldbart2003, ggm_Shimony1995Apr}. It is defined as an optimized distance of the given state from the set of all  non-genuinely multiparty entangled states, given by
\begin{equation}\mathcal{E}(\ket{\psi_{N}}) = 1 -\Lambda_{max}^{2}(\ket{\psi_{N}}) \label{eq:ggm_definition}\end{equation}
\noindent
where $\mathrm{\Lambda_{max}(\ket{\psi_{N}}) = max |  \langle \chi \mid \psi_{N} \rangle |}$   with the maximization being over all $\ket{\chi}$ that are not genuinely multiparty entangled.
An equivalent form of the above equation is \cite{sen_pra_2010}
\begin{equation}\mathcal{E}(|\psi_N\rangle) = 1 - \max \left\{ \lambda^2_{I : L} | I \cup L = \{A_1,...,A_N\}, I \cap L = \varnothing \right\}\end{equation}
 where $\lambda_{I : L}$ is the maximal Schmidt coefficient in the bipartite
split \(I : L\) of $\ket{\psi_{N}}$.

{\it Logarithmic negativity.} On the other hand, bipartite entanglement in two-qudit systems, \(\rho_{ij}\) between two nodes, \(i\) and \(j\),  \cite{horodecki_pla_1996,peres_prl_1996,vidal_pra_2002} is defined as $$\mathcal{LN}(\rho_{ij}) = \log_2(2\mathcal{N}(\rho_{ij})+1),$$ where negativity, denoted as $\mathcal{N}$, is the absolute sum of all the negative eigenvalues of the partially transposed state ($\rho_{ij}^{T_i}$ or $\rho_{ij}^{T_j}$) with the partial transposition being taken with respect to either \(i\) or \(j\). Notice that when biaprtiite entanglement is computed in any two pairs of subsystems in the different stages of the HHL algorithm used for arbitrary linear systems of equations, each subsystems are, in general, higher dimensional systems, and hence nonvanishing \(\mathcal{LN}\)  implies that the entanglement present in the system is distillable \cite{horodeckidistillable}.

{\it Coherence.}
As we know measure of coherence is basis dependent, 
$l_1$ norm of coherence \cite{baumgratz_prl_2014,streltsov_rmp_2017} in the computational basis  is defined as \(   \mathcal{C}'(\rho)= \sum_{i \neq j}\abs{\rho_{ij}}\), where \(\rho_{ij}\) are the elements of the density matrix \(\rho\). The coherence of maximally coherent state in the Hilbert space of dimension $D$ under this measure is $D-1$. In order to make the coherence measure dimension-independent,  we scale the above measure by $D-1$ and hence the coherence measure computed in this work reads as \(\mathcal{C}(\rho) = \frac{\sum_{i \neq j}\abs{\rho_{ij}}}{D-1}\).

Let us first address the question about genuine multipartite entanglement (a $N$-party pure state is genuinely multipartite entangled if it is not a product across any bipartition) - ``In the HHL algorithm, is genuine multipartite entanglement at all required?". We answer the question affirmatively. 

We first note that GGM is vanishing at the first and third steps, given by $|\psi_1\rangle$ and $|\psi_3 \rangle$ respectively since they are product across at least one of the biparitions of the tripartite state shared between $\mathbf{\Lambda}$, $\mathbf{U}$, and $\mathbf{R}$. On the other hand, let us now show that nonvanishing GME in the second step is necessary for the success of the HHL algorithm. In the next section, we will confirm this by computing GGM and its trends with the system parameters. 

{\bf Theorem 1.} For a successful run of the HHL algorithm, genuine multipartite entanglement should be nonvanishing in the second step.
\begin{proof}
To establish the genuine multipartite entanglement present between three parties in the second step, we prove that all the single-site local density matrices, $\rho_{\mathbf{\Lambda}}$, $\rho_\mathbf{U}$ and $\rho_{\mathbf{R}}$ of $|\psi_2\rangle$ are mixed, i.e.,  $\Tr(\rho_X^2) < 1$, for $X = \mathbf{\Lambda}, \mathbf{U}, \text { and } \mathbf{R}$. Since the total state is pure, mixed local density matrices ensure that no biparition is product, thereby proving nonvanishing GME of $|\psi_2\rangle$.  

For \(N\)  number of linear equations, the state in the second step is represented in Eq. (\ref{eq:step2}). Since each \(\ket{\lambda_i}_{\mathbf{\Lambda}}\) and \(\ket{u_i}_{\mathbf{U}}\) are orthogonal to each other, the reduced density matrices of $\mathbf{\Lambda}$ and $\mathbf{U}$ are given by \(\rho_{\mathbf{\Lambda}} = \text{Tr}_{\mathbf{UR}} (\ket{\psi_2}\bra{\psi_2}) = \sum_{i=1}^{N} \beta_i^2 \ket{\lambda_i}\bra{\lambda_i}\)  and \(\rho_{\mathbf{U}} = \text{Tr}_{\mathbf{\Lambda R}} (\ket{\psi_2}\bra{\psi_2}) =  \sum_{i=1}^{N} \beta_i^2 \ket{u_i}\bra{u_i}\) which lead to $ \tr(\rho_{\mathbf{\Lambda}}^2)=\tr(\rho_{\mathbf{U}}^2) = \sum_{i=1}^{N}\beta_i^4 =1- 2 \sum_{i<j=1;i\neq j}^{N} \beta_i^2\beta_j^2 \leq 1$ ( since $ \sum_{i=1}^{N} \beta_i^2 = 1$). The equality holds only if any of the \(\beta_i=1\) and the rest of them vanish, which implies that the constant vector, \(\vec{b}\) is an eigenvector of the matrix \(\mathbf{A}\). It indicates that in a nontrivial situation,  nonvanishing entanglement of $|\psi_2\rangle$ is present in the bipartitions   $\mathbf{\Lambda:UR}$ and $\mathbf{U: \Lambda R}$. 

On the other hand, the reduced density matrix of the auxiliary qubit is not diagonal and reads as 
\begin{align}
\rho_{\mathbf{R}} = & \begin{bmatrix}
1-\sum_{i=1}^{N}a_i  & \sum_{i=1}^{N} b_i  \\
\sum_{i=1}^{N}b_i   & \sum_{i=1}^{N}a_i 
\end{bmatrix}, \nonumber\\
\label{eq:rho_R}
\end{align}
where $a_i = \beta_i^2\frac{C^{2}}{\lambda_{i}^{2}} $  and $b_i = \beta_i^2\sqrt{1-\frac{C^{2}}{\lambda_{i}^{2}}} \cdot \frac{C}{\lambda_{i}}$. Therefore, the purity of the state is given by  $\tr(\rho_{\mathbf{R}}^2) =  1- 2(\sum_{i=1}^{N}a_i - \sum_{i,j=1}^{N}a_i a_j  -\sum_{i,j=1}^{N}b_i b_j ). $
In order to prove that the state is a mixed state, we use \(\det(\rho_{\mathbf{R}})\ge 0\) which gives 
$\sum_{i=1}^{N}a_i - \sum_{i,j=1}^{N}a_i a_j  - \sum_{i,j=1}^{N}b_i b_j \geq 0 $, 
and hence
$ \tr(\rho_\mathbf{R}^2) =  1- 2(\sum_{i=1}^{N}a_i - \sum_{i,j=1}^{N}a_i a_j  -\sum_{i,j=1}^{N}b_i b_j ) \leq 1. $ The equality occurs exactly when one of the \(\beta_i\) is unity, thereby leading to the trivial situation. 
 Hence the proof.   
\end{proof}

Let us now scrutinize how the bipartite entanglement is distributed in this protocol. Unlike GME, bipartite entanglement between subsystems are always present in different stages of the algorithm. In the first step,   the reduced bipartite state, $\rho_{\mathbf{\Lambda U}}$, is entangled although $\ket{\psi_1}$ has vanishing GGM. Similarly, in the last step, instead of $\mathbf{\Lambda}$ and $\mathbf{U}$, $\rho_{\mathbf{UR}}$ possesses nonvanishing bipartite entanglement. 

We now investigate closely the status of two-party entanglement in the second step of the algorithm since in this case, the state has nonvanishing GME and so their reduced states cannot be trivially shown to be bipartite entangled.

{\bf Theorem 2.} Bipartite entanglement between subsystems $\mathbf{\Lambda}$ and $\mathbf{U}$ parties is nonvanishing while it vanishes for the other two pairs in the second step of the HHL algorithm.  

\begin{proof} 
To show nonvanishing bipartite entanglement, we compute logarithmic negativity $(\mathcal{LN})$ of $\rho_{\mathbf{\Lambda U}}$, $\rho_{\mathbf{RU}}$ and $\rho_{\mathbf{\Lambda R}}$ \cite{vidal_logneg_2002} of \(|\psi_2\rangle\). Let us first consider two-party reduced density matrix between $\mathbf{U}$ and $\mathbf{R}$, which can be written as 
\begin{align*}
\rho_{\mathbf{UR}} = \sum_{i=1}^{N}\beta_i^2 &\ket{u_i}\left(\sqrt{1-\frac{C^{2}}{\lambda_{i}^{2}}}\ket{0} + \frac{C}{\lambda_{i}}\ket{1}\right)\\ & \bra{u_i}  \left(\sqrt{1-\frac{C^{2}}{\lambda_{i}^{2}}}\bra{0} + \frac{C}{\lambda_{i}}\bra{1}\right).
\end{align*}
In the basis $\{ |u_i 0 \rangle, |u_i 1 \rangle \}$,  $ \forall i \in \{1, 2, \ldots N\}$, it can be written as $\rho_{\mathbf{UR}} = \bigoplus_{i=1}^N \mathbf{\omega_i} $ 
where
\[ \mathbf{\omega_i} = \begin{bmatrix}
 \beta_i^2 \left( 1- \frac{C^2}{\lambda_i^2} \right) & \beta_i^2 \sqrt{1 - \frac{C^2}{\lambda_i^2}} \cdot \frac{C}{\lambda_i} \\
 \beta_i^2 \sqrt{1 - \frac{C^2}{\lambda_i^2}} \cdot \frac{C}{\lambda_i} & \beta_i^2 \frac{C^2}{\lambda_i^2} \\
\end{bmatrix}. \]
It is evident that after partial transposition over the second party,  $\mathbf{R}$, the state  does not change, i.e., $ (\rho_{UR})^{T_R} = \rho_{UR} $. Hence, there is no bipartite (distillable) entanglement in $\rho_{\mathbf{UR}}$ since $\rho_{\mathbf{UR}}$ is in higher dimensional space \cite{horodeckidistillable, horodecki_ent_distillable_1999}. 
Similarly, $\rho_{\mathbf{\Lambda R}}$ in the basis $ \{ |\lambda_i 0 \rangle, |\lambda_i 1 \rangle \} $ $\forall i \in \{1, 2 \ldots N \}$ can be written as $\bigoplus_{i=1}^{N} \mathbf{\omega_i^\prime} $
with $\mathbf{\omega_i^\prime} = \mathbf{\omega_i}$, which again implies  vanishing $\mathcal{LN}$, thereby vanishing bipartite (distillable) entanglement. 

After tracing out the auxiliary qubit $\mathbf{R}$ and by performing partial transposition over $\mathbf{U}$ of $\rho_{\mathbf{\Lambda U}}$, we obtain  
$ (\rho_{\mathbf{\Lambda U}})^{T_\mathbf{U}} = 
\begin{bmatrix}
 \mathcal{B}  & 0  \\
0 & \tilde{\mathcal{B}} \\
\end{bmatrix}.
$ Here $\mathcal{B}$ = diag$\{\beta_0^2, \beta_1^2 \ldots,\beta_N^2 \} $ and $ \tilde{\mathcal{B}} = \bigoplus_{i , j = 1;i \neq j}^{N}  \zeta_{ij} $, such that the dimension of \(\tilde{\mathcal{B}}\) is \(\binom{N}{2} \times \binom{N}{2} \) containing \(2\times 2\) block as
\[ \zeta_{ij;i\neq j} = 
\begin{bmatrix}
0 & \beta_i\beta_j(\lambda_i' \lambda_j' + \tilde{\lambda_i} \tilde{\lambda_j})  \\

\beta_i\beta_j(\lambda_i' \lambda_j' + \tilde{\lambda_i} \tilde{\lambda_j}) &  0 \\
\end{bmatrix}.
\]
Here $\tilde{\lambda_a} = \frac{C}{\lambda_a}$ and  $\lambda_a' = \sqrt{1-\frac{C^2}{\lambda_a^2}}$. This matrix has $N$ number of negative eigenvalues and their absolute sum \cite{vidal_logneg_2002} is given by
\[ \mathcal{N}(\rho_{\mathbf{\Lambda U}}) = \sum_{i,j =1;i<j}^{N} |\beta_i\beta_j|\Big(\sqrt{1-\frac{C^2}{\lambda_i^2}}\sqrt{1-\frac{C^2}{\lambda_j^2}} + \frac{C}{\lambda_i} \frac{C}{\lambda_j}\Big),\]
which is strictly positive when none of the \(\beta_i\)s vanish. 
Therefore, the reduced density matrix without the auxiliary system is entangled. 

\end{proof}
\textit{Remark 1.} We first notice that the tripartite state, \(|\psi_2\rangle\), is not symmetric under the permutations of the parties. Moreover, from the structures of the single-site local density matrices and the bipartite entanglement studied in Theorem $1$ and $2$, it is evident that the multipartite state cannot possess maximal GGM  although nonvanishing GME is required to have nontrivial computation in the HHL algorithm.

\subsubsection*{Relation between quantum coherence and success probability} 

As we have shown,  entanglements of different types are crucial for the nontrivial solutions of the HHL algorithm. It is always interesting to probe whether any other form of nonclassicality is also present during the execution of the algorithm. 
In this respect, one of the potential candidates to probe is quantum coherence \cite{streltsov_rmp_2017}. We indeed find that nonvanishing quantum coherence measured by $l_1$-norm  \cite{baumgratz_prl_2014,streltsov_rmp_2017}, is present when the tripartite  states as a whole in all the steps are considered and when single site reduced density matrices,   $\rho_\mathbf{U}$ or $\rho_\mathbf{R}$ is determined. 


{\bf Proposition 1.} The $l_1$-norm coherences of the global and some of the single-site reduced density matrices are nonvanishing. 


\begin{proof}

Since  $\rho_{\Lambda}$ is diagonal in the computational basis, coherences in all the steps vanish. On the other hand, \(\rho_\mathbf{U} =  \sum_{i=1}^{2} \beta_i^2 \ket{u_i}\bra{u_i}\) which is diagonal in the basis of \(\ket{u_i}\) although the state has nonzero coherence in the computational basis, both the second and the third steps. 

From Eq. (\ref{eq:rho_R}), $\rho_\mathbf{R}$ written in the computational basis has nonvanishing off-diagonal terms which lead to nonvanishing $l_1$-norm coherence given by 
\begin{equation}
    \mathcal{C}_\mathbf{R} = 2\sum_{i=1}^{N}\beta_i^2\sqrt{1-\frac{C^{2}}{\lambda_{i}^{2}}}\cdot \frac{C}{\lambda_{i}},
    \label{eq:coherence_rho_R}
\end{equation}
which again vanishes for the trivial situation. Similar situation emerges also in the last step by considering \(|\psi_3\rangle\). Due to the presence of nonvanishing coherence in some of the subsystems, \(\mathcal{C}_{\mathbf{\Lambda U R}} \neq 0\) in all the steps of the HHL algorithm. 
\end{proof}

\textit{Remark 2. } Comparing Eq. (\ref{eq:coherence_rho_R}) with $\mathcal{SP}$ in Eq. (\ref{eq:soln}), one realizes that $C_\mathbf{R}$ is connected with $\mathcal{SP}$. In the next section, we will make the connection of the success probability with quantum features like entanglement and quantum coherences of the resulting state in each step  more concrete. 

\begin{figure}
\includegraphics[width=1\linewidth]{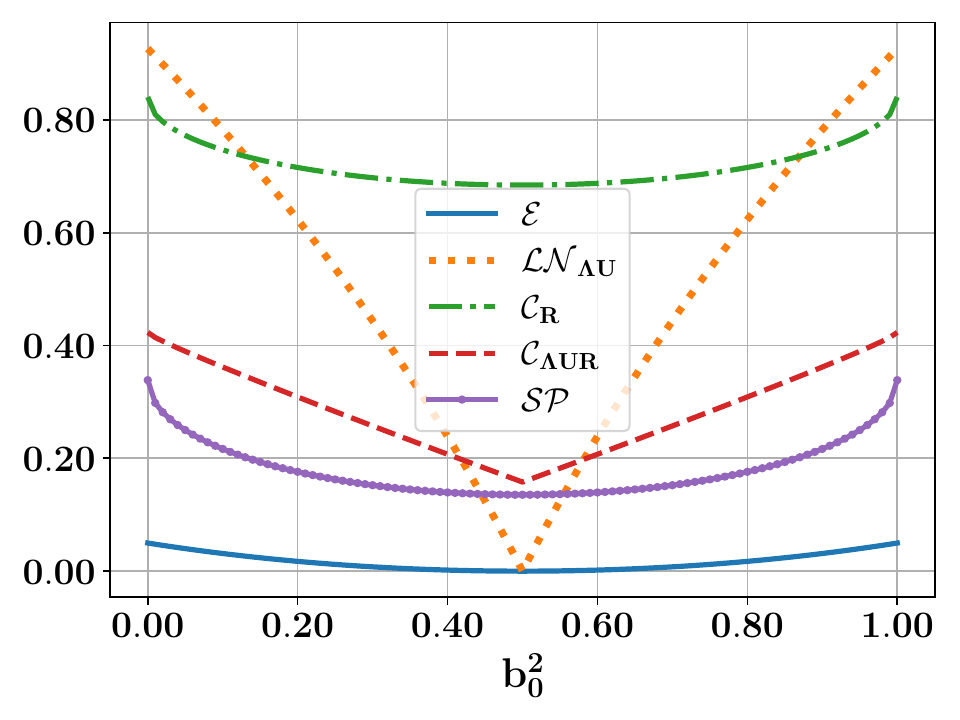}
\caption{ \textbf{Variation of different nonclassical correlation measures ($\mathcal{E}$, $\mathcal{LN}_{\mathbf{\Lambda U}}$, $\mathcal{C}_{R}$, $\mathcal{C}_{\mathbf{\Lambda U R}}$) and success probability $\mathcal{SP}$  (ordinate) with respect to $b_0^2$ (abscissa) in the second step}. This figure shows the generic behavior for different nonclassical correlation measures  of $\ket{\psi_2}$ for a set two linear equations given in Eq. (\ref{eq:2d}). Since the vector $\vec{b}$ is normalized, we only have one parameter to vary.
All the axes are dimensionless.}
\label{fig:ggm_b0}
\end{figure}

\section{Connecting quantum resources with  computation process}
\label{sec:connect}

Up to now, we have shown the necessity of quantum resources for obtaining nontrivial results in the HHL algorithm which we establish by proving the nonvanishing quantum correlations or quantum coherence of the states produced during the execution of the algorithm. We will now concentrate on their trends with the variations of the parameters involved. Towards this quantitative analysis, we illustrate with some examples in different dimensions.


{\it Two-dimensional linear systems of equations.} Let us start with an example in two dimensions which was used to demonstrate HHL algorithms in laboratories \cite{pan_pra_2014} given as  
\begin{equation}
\mathbf{A} = \frac{1}{2} \begin{pmatrix}
        3 &  1\\
        1 & 3
    \end{pmatrix},  \hspace{5mm} \text{and} \,\, \vec{b} = \begin{pmatrix}
        b_{0}\\
        b_{1}
    \end{pmatrix},
    \label{eq:2d}
\end{equation}
where $b_{0}^{2} + b_{1}^2 = 1$. The corresponding eigenvalues and eigenvectors of the matrix \(\mathbf{A}\) are 
\begin{eqnarray}
\nonumber \lambda_{1} &=& 1, \hspace{5mm} \ket{u_{1}} = \frac{1}{\sqrt{2}} ( \ket{0} - \ket{1}),\\
\text{and} \, \, \lambda_{2} &=& 2, \hspace{5mm} \ket{u_{2}} = \frac{1}{\sqrt{2}} ( \ket{0} + \ket{1}).
\end{eqnarray} 
As discussed before, GGM of $\ket{\psi_1}$ and $\ket{\psi_3}$ vanish. For calculating GGM of $\ket{\psi_2}$, let us write the register qubits, \(\ket{b}_{\mathbf{U}}\) in a linear combination of eigenvectors of \(\mathbf{A}\) with \(\beta_1 = \frac{1}{\sqrt{2}} (b_0-b_1)\) and \(\beta_2 = \frac{1}{\sqrt{2}} (b_0 + b_1)\). It is followed by a controlled rotation with the auxiliary qubit being rotated  with an angle, $\theta =2 \sin^{-1} \frac{C}{\mathbf{\lambda}}$. Here we choose $C = 0.736$ although any other  values of \(0.5\leq C \leq 1\) leads to a qualitatively similar result since it is of the order of \(\mathcal{O}(\frac{\lambda_1}{\lambda_2})\). The GGM of $\ket{\psi_2}$ vanishes only when $b_0^2=\frac{1}{2}$ which is a trivial situation.  Otherwise, GGM increases (with the variation of $b_0^2$) symmetrically around $b_0^2=\frac{1}{2}$ as depicted in Fig. \ref{fig:ggm_b0}. Similar characteristic emerges also for $\mathcal{LN}$ of $\rho_{\mathbf{\Lambda U}}$, denoted by $\mathcal{LN}_{\mathbf{\Lambda U}}$, although the sharp kink present in $\mathcal{LN}_{\mathbf{\Lambda U}}$ is missing for GGM. Both normalized coherence for the entire state and the register qubit behave similarly with $b_0^2$ like GGM and $\mathcal{LN}_{\mathbf{\Lambda U}}$. We observe that both $C_{\mathbf{\Lambda U R}}$ and $C_\mathbf{R}$ are nonvanishing for alll values of $b_0^2$ although both of them are minimum at  $b_0^2=\frac{1}{2}$. Interestingly, the pattern of  $C_{\mathbf{R}}$ matches with GGM, especially  near \(b_0^2 = \frac{1}{2}\) while $\mathcal{LN}_{\mathbf{\Lambda U}}$  and $C_{\mathbf{\Lambda U R}}$ diverges at $b_0^2 = \frac{1}{2}$. Notice, further, that the pattern of $C_{\mathbf{R}}$ is exactly similar to the success probability, \(\mathcal{SP}\), comparing them in Fig. \ref{fig:ggm_b0}.

{\it Example of a three-dimensional linear systems of equations.} We take an example of a Hermitian \(3\times 3\) matrix, given by 
\begin{equation}
\frac{1}{6} \begin{bmatrix}
    14 & -4 & -4 \\
    -4 & 11 & -1 \\
    -4 & -1 & 11 \\
\end{bmatrix}
\begin{bmatrix}
x_0 \\
x_1 \\
x_2
\end{bmatrix}
= \begin{bmatrix}
b_0 \\
b_1 \\
b_2
\end{bmatrix}.
\label{eq:3d}
\end{equation}
The eigenvalues and the eigenvectors of the corresponding matrix are given by
\begin{eqnarray}
\nonumber \lambda_{1} &=& 1 , \hspace{5mm} \ket{u_{1}} = \frac{1}{\sqrt{3}} ( \ket{0} +\ket{1} + \ket{2}),\nonumber\\
\lambda_{2} &=& 2, \hspace{5mm} \ket{u_{2}} = \frac{1}{\sqrt{2}} ( \ket{1} - \ket{2}),\\
\text{and}\,\, \lambda_{3} &=& 3, \hspace{5mm} \ket{u_{3}} = \frac{1}{\sqrt{6}} ( -2\ket{0} + \ket{1} +\ket{2}). \nonumber
\end{eqnarray}
The constant vectors are represented as a linear combination of the eigenvectors of \(\mathbf{A}\) which are \(\beta_1 = \frac{1}{\sqrt{3}} (b_0 + b_1 + b_2) \), \(\beta_2 = \frac{1}{\sqrt{2}} (b_1 - b_2)\) and \(\beta_3 = \frac{1}{\sqrt{6}} (-2b_0 + b_1 +b_2)\). In order to determine the role of GGM, we only focus on the state \(\ket{\psi_2}\), since it  only contains nontrivial GME. 

\begin{figure}
\includegraphics[width=1\linewidth]{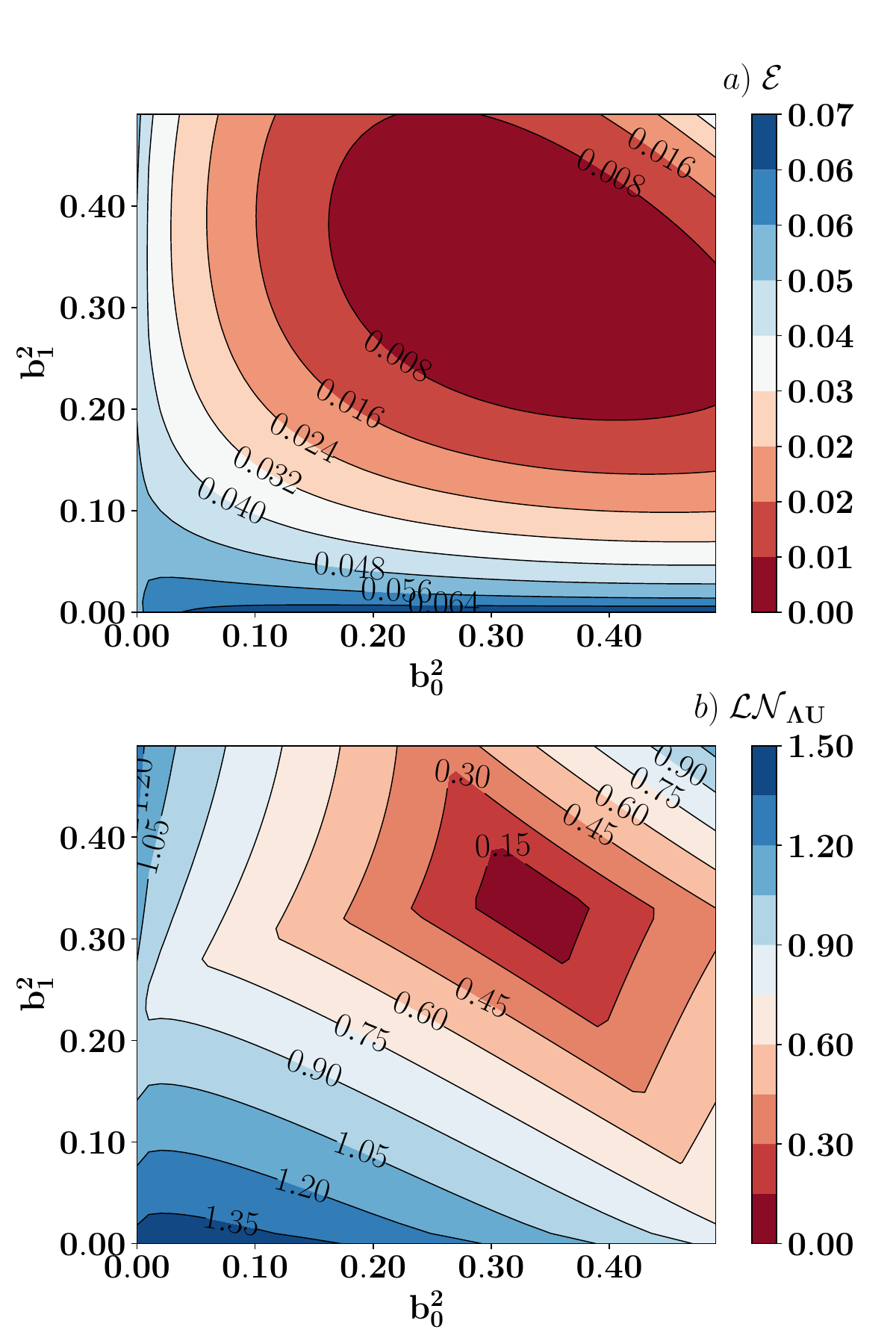}
\caption{ \textbf{Contour plots of  GGM ($\mathcal{E}$) (upper panel) and logarithmic negativity ($\mathcal{LN}_{\mathbf{\Lambda U}}$) (lower panel) for  three-dimensional linear systems of equations.} The variation of $\mathcal{E}$ (ordinate) and $\mathcal{LN}_{\mathbf{\Lambda U}}$ (ordinate) as functions of $b_{0}^{2}$ and $b_{1}^{2}$ (abscissa) of \(|\psi_2\rangle\) in Eq. (\ref{eq:3d}). Note that at points where the vector $\vec{b}$ is proportional to the eigenvalues of the matrix $\mathbf{A}$, both entanglement measures, $\mathcal{E}$ and $\mathcal{LN}_{\mathbf{\Lambda U}}$, vanish. These points correspond to trivial instances where the input state is the same as the output state. Comparing upper and lower panels, it is interesting to note that GME content present in \(\ket{\psi_2}\) is much smaller than the bipartite entanglement in subsystems.   All the axes are dimensionless. }
\label{fig:ggm_3x3}
\end{figure}

\begin{figure}
\includegraphics[width=1\linewidth]{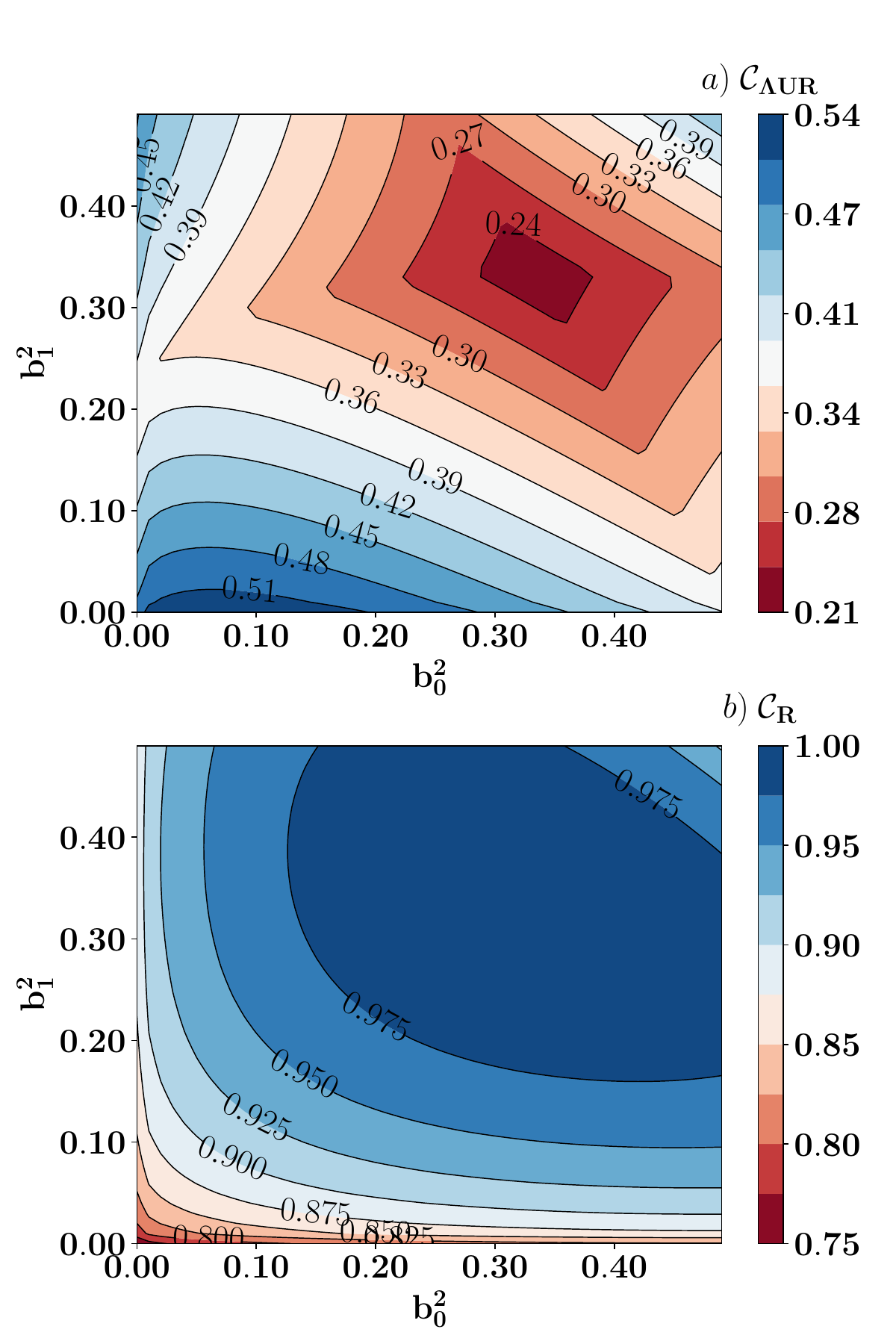} 
\caption{\textbf{Contour plots of coherences of  $\rho_{\mathbf{\Lambda U R}}$ (upper panel) and  $\rho_\mathbf{R}$ (lower panel)  against $b_0^2$ and $b_1^2$ of \(\ket{\psi_2}\) in Eq. (\ref{eq:3d}).} We observe that the behavior of coherence for these two states is opposite to each other. The coherence of the entire state follows a similar pattern to $\mathcal{E}$ and $\mathcal{LN}_{\mathbf{\Lambda U}}$ for all dimensions, whereas the coherence of the state $\rho_R$ which is same as in the second and third steps is connected  with the success probability of the system. All the axes are dimensionless.}
\label{fig:coherence_3d} 
\end{figure}

In this case, there are two unknown parameters, $b_0^2$ and $b_1^2$, which can be varied to study \(\mathcal{E}\), $\mathcal{LN}_{\mathbf{\Lambda U}}$, $C_{\mathbf{\Lambda U R}}$ and $C_\mathbf{R}$. When $b_0^2 = b_1^2 = b_0^2 = \frac{1}{3}$, GGM and $\mathcal{LN}_{\mathbf{\Lambda U}}$ both vanish which represent the trivial solution. Comparing Fig. \ref{fig:ggm_b0} and contours of Fig. \ref{fig:ggm_3x3}, we observe that the qualitative behaviors of $\mathcal{E}$ and $\mathcal{LN}_{\mathbf{\Lambda U}}$ are same in both the two- and three-dimensional linear systems of equations, and similar behavior also emerges for coherence, as depicted in Fig. \ref{fig:coherence_3d}. Hence, we can possibly conclude that with the increase of the dimensions of linear systems, the roles of entanglement and coherence in the HHL algorithm remain the same as also seen in the proofs of the preceding section. 

\begin{figure}
\includegraphics[width=\linewidth]{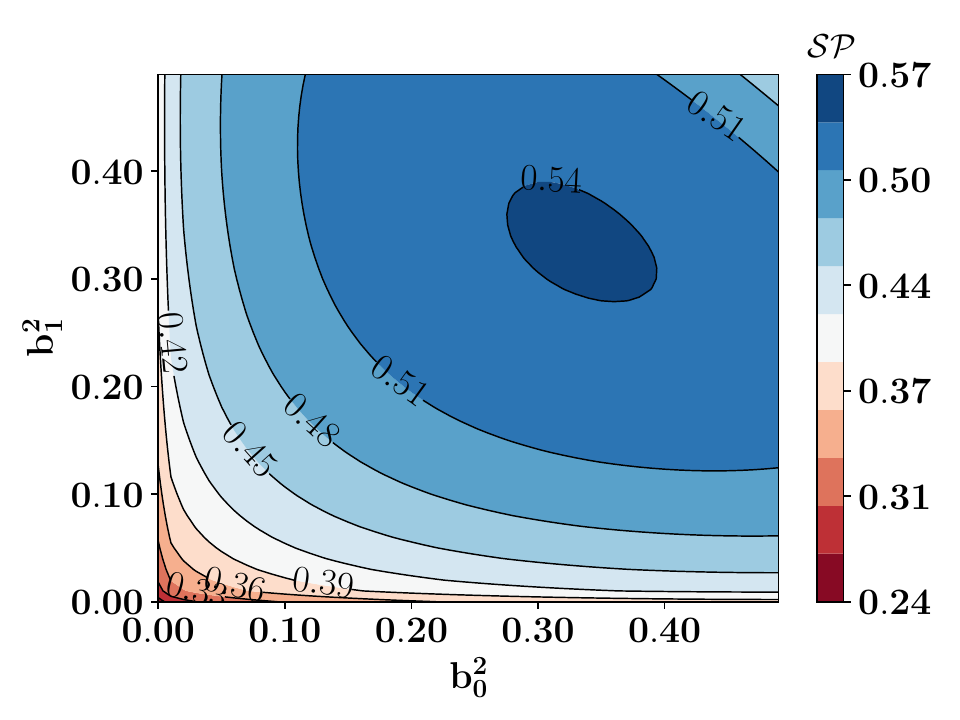}
\caption{ \textbf{Contour plot of success probability $\mathcal{SP}$ (\(z\)-axis) as a function of  $b_0^2$ (\(x\)-axis) and $b_1^2$ (\(y\)-axis) in Eq. (\ref{eq:3d}).} The pattern of the success probability in the read-out is similar to  $\mathcal{C}_R$ in the second and third steps as shown in Fig. \ref{fig:coherence_3d}. All the axes are dimensionless. }
\label{fig:sp_3d}
\end{figure}

Another important point to stress here is that although GGM content in the state is low in the second step and vanishing in other steps, the bipartite entanglement content, $\mathcal{LN}_{\mathbf{\Lambda U}}$ between $\mathbf{\Lambda}$ and $\mathbf{U}$ is moderately  high. We also notice that a significant amount of coherences, especially in \(\rho_{\mathbf{R}}\) are also present in the second step. It is important to notice that the success probability depicted in Fig. \ref{fig:sp_3d} resembles with that of coherence in the register qubit, $\mathbf{R}$ as seen in Fig. \ref{fig:coherence_3d}(b). 

Let us summarize two interesting aspects which are manifested from the above examples. (1) Quantumness shared between  $\mathbf{U}$ and $\mathbf{R}$  or present in a party $\mathbf{R}$ is comparatively higher than the other parts. Notice that $\mathbf{R}$ is the register qubit used for readout, thereby responsible for the successful implementation of the protocol. Hence, one can argue that at the end,  site \(\mathbf{R}\) contributes maximally to the scheme. (2) Although nonvanishing quantum features are required for the nontrivial computation of the HHL algorithm, the quantitative analysis of quantum correlations possibly indicate that the high genuine tripartite nonclassicality is not so essential. We will confirm such intuition by connecting quantum properties with the complexity of the algorithm.

\subsection{Complexity vs Nonclassical Resources}

We have already established that entanglement is a necessary condition for the algorithm to proceed. Let us analyze the variation of entanglement measures with the performance indicator, $\kappa$ which is the ratio between the highest and the lowest eigenvalue of the matrix $\mathbf{A}$ \cite{HHL_PRL_2009}. The runtime of the algorithm scales as $\kappa^2 \log N/ \epsilon$ with $\epsilon$ being the error in the output state, i.e., the higher the $\kappa$ the worse the time complexity. The quantum advantage is achieved when both \(\kappa\) and \(\frac{1}{\epsilon}\) behaves as \(\text{poly} \log N\). 

\begin{figure}
\includegraphics[width=\linewidth]{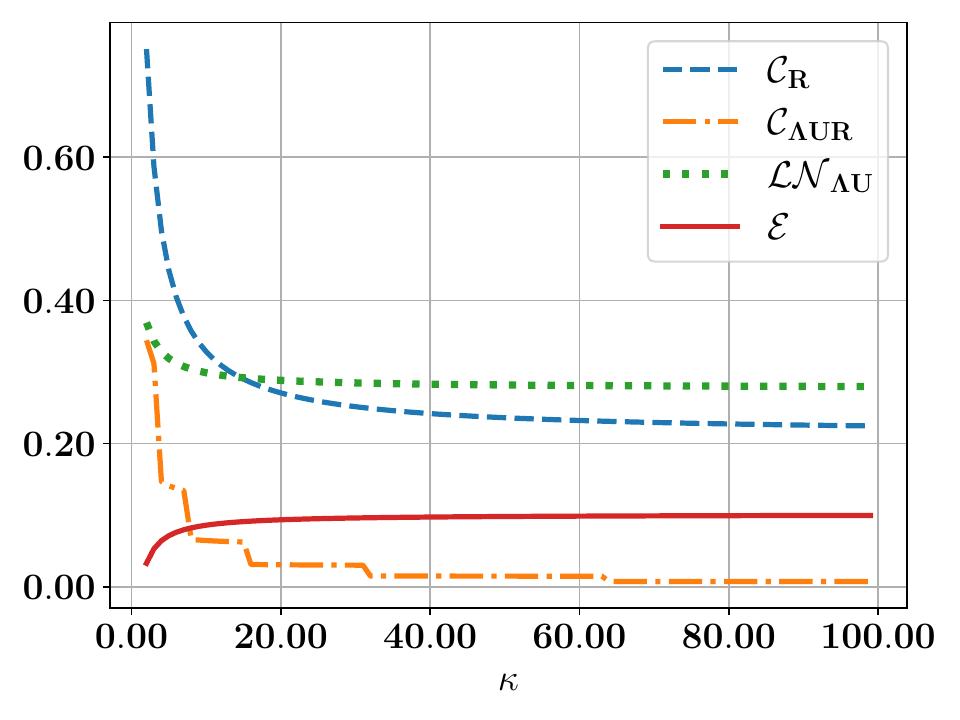}
\caption{\textbf{ Nonclassical correlation measures (ordinate) vs time complexity, $\kappa$ (abscissa) for and two-dimensional linear systems in Eq.  (\ref{eq:2d}).}  We take $b_{0}=0.3$ .  As in two-dimensional linear system, we observe that $\mathcal{E}$ increases with $\kappa$ while the coherence ($C_\mathbf{R}$ and $C_{\mathbf{\Lambda UR}}$) and logarithmic negativity ($\mathcal{LN}_{\mathbf{\Lambda U}}$ ) decreases with $\kappa$. It demonstrates that the increase in $\mathcal{E}$ can create  hindrances in the execution of the algorithm while the increase in $\mathcal{C}$ and $\mathcal{LN}$ are associated with more efficient functioning of the algorithm. All the axes are dimensionless.}
\label{fig:kappa_3x3}
\end{figure}

Considering both the examples of two- and three-dimensional linear systems of equations, we interestingly observe that $\kappa$ increases with the increase of the genuine multipartite entanglement content as shown in Fig. \ref{fig:kappa_3x3}. It implies that although GME is required for the nontrivial solution of the algorithm, higher amount of genuine quantum correlation can create hindrances, thereby worsening its performance quantifiers. Similar results are recently obtained for other quantum algorithms\cite{grover_entanglement_PRA_2013, multi_ent_grover_2012, entanglement_bernstein_pra_2022}. However, both  \(\mathcal{LN}_{\mathbf{\Lambda U}}\) in \(|\psi_2\rangle\) and coherences,  \(\mathcal{C}_{\mathbf{\Lambda U R}}\) and  \(\mathcal{C}_{\mathbf{R}}\), decreases with the increase of \(\kappa\) (see Fig. \ref{fig:kappa_3x3}).  
Summarizing, in the HHL algorithm, nontrivial execution of the algorithm requires high amount of quantum features in local subsystems  although high multipartite quantum correlations can be an obstacle for the implementation.

\section{Effects of Imperfections on HHL algorithm}
\label{sec:disorder}
Imperfection or noise has eminent presence in all types of devices, thereby affecting the outputs of any circuit. In this respect, it is natural to assume that any unitary gate involved in the protocol can only be executed with some errors. One may expect that the performance of any devices gets degraded in presence of imperfections or noisy environment. Counter-intuitively, it was reported that disorder can sometimes enhance certain properties like magnetization, entanglement in the system \cite{garnerone_prl_2009,jacobson_prb_2009,niederberger_pra_2010,sadhukhan_disorder_2016} and phenomena like localization \cite{bloch_rmp_2019}, violation of eigenstate thermalization hypothesis \cite{anatoli_adv_2016} etc. At the same time, it was also highlighted that quantum devices like quantum thermal machines \cite{konar_pra_2022}, quantum networks \cite{halder_pra_2022}, quantum communication protocols \cite{muhuri_arxiv_2022}, quantum metrology \cite{bhattacharyya_arxiv_2023,monika_arxiv_2023} may not always get affected by the imperfections in the operations involved in the schemes. 

\begin{figure}
\includegraphics[width=\linewidth]{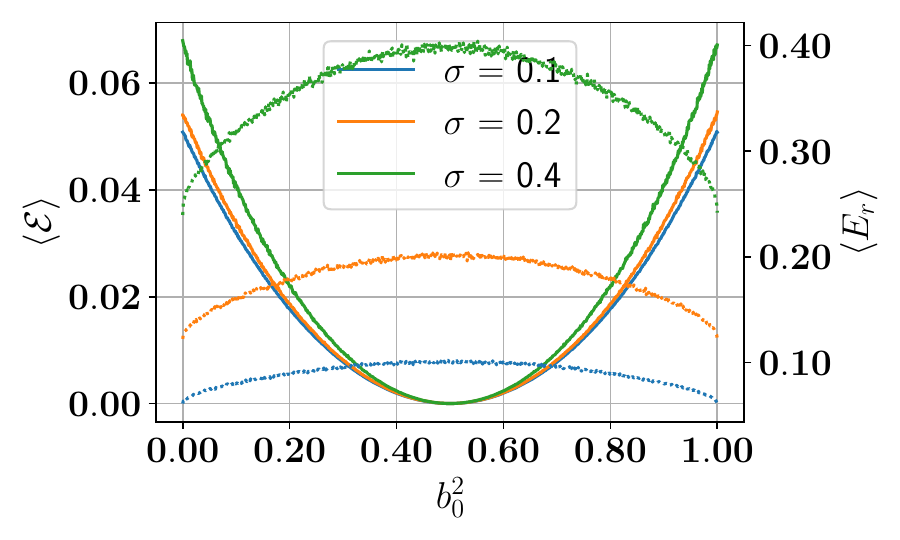}
\caption{ \textbf{Quenched average GGM ($\expval{\mathcal{E}}$) and error ($\expval{E_r}$) (ordinate) against  $b_{0}^{2}$ (abscissa) for the two-dimensional linear systems of equation (Eq. (\ref{eq:2d})).}
As we introduce imperfection in the algorithm in the form of Gaussian disorder, $\frac{\tilde{\theta_i}}{2} = \frac{C}{\lambda_i} + \epsilon_i$, with \(\epsilon_i\) being chosen from gaussian distribution with vanishing mean and standard deviation,  \(\sigma\),   both the quenched average GGM in the second step, $\expval{\mathcal{E}}$ (solid line),  and  $ \expval{E_r}$ (dotted line) accumulated at the end of the algorithm increase with the increase of disorder, \(\sigma\). For obtaining quenched average quantities, averaging is performed over \(10^4\) realizations for each $b_0^2$.  This establishes that while $\mathcal{E}$ is a necessary condition, it is not favorable for the execution of the algorithm. All the axes are dimensionless.}
\label{fig:ggm_error_disorder}
\end{figure}

In our work, we introduce imperfections in the gates involved in the HHL algorithm. Although this algorithm has already been realized in experimental circuits \cite{pan_pra_2014}, the consequence of disorder on its performance is not clear yet. 
Let us consider the scenario when the imperfection is present in the controlled rotation involved in the HHL algorithm. In other words, controlled rotation contains some errors which are chosen randomly from some probability distribution. 
Both quantum resources and the errors accumulated in this scheme is computed by averaging over random sample choices of the parameters involved, known as quenched averaging. Mathematically, the quenched average quantity \(\mathcal{Q}\) is defined as
\begin{equation}    \left\langle\mathcal{Q}\left(\langle g \rangle, \sigma_{g}\right)\right\rangle = \int \mathcal{P}(g)\mathcal{Q}(g)d(g),
\label{eq:quenched}
\end{equation}
where $g$ is the parameter values, chosen from a distribution ($\mathcal{P}(g)$) of mean $\langle g \rangle$ and standard deviation $\sigma_g$  quantifying the strength of the disorder. In our case, the probability distribution is chosen to be Gaussian. 

Let us illustrate the situation by considering two-dimensional linear system of equations as discussed in the preceding section. Conditional rotation on ancillary qubits is given by \(\theta_i=2\sin^{-1}(\frac{C}{\lambda_i})\) with \(\lambda_i\) being the eigenvalue of matrix \(\textbf{A}\). An imperfect rotation can be represented as
\begin{equation}
    \frac{\tilde{\theta_i}}{2}= \sin^{-1}(\frac{C}{\lambda_{i}}) + \epsilon_{i},
\end{equation}
where \(\epsilon_i\) is the amount of imperfection in the rotation. We choose \(\epsilon_i\) randomly from a Gaussian distribution with mean zero and standard deviation \(\sigma\). We observe the effect of disorder on the first example of two linear equations, although  we also check numerically the effects on other sets of linear equations. First, we try to understand the impact of the disorder analytically on the solution of the state \(\ket{\psi_x}\). Assume \(\theta_i\)'s to be small, so that we can make the approximation that $\frac{\theta_i}{2} \approx \frac{C}{\lambda_i}$ which can be implemented by tuning the circuit constant \(C\). Therefore, $\frac{\tilde{\theta_i}}{2} = \frac{C}{\lambda_i} + \epsilon_i$ and after substituting $\tilde{\theta_i}$ in place of $\theta_i$ in  Eq. (\ref{eq:soln}), we get
\begin{align}
\ket{{\Psi}_x} &=  \frac{1}{2}\left (
    \begin{aligned}
        & (b_{0}-b_{1}) \frac{C}{\lambda_{1}} +  (b_{0}-b_{1}) \epsilon_1\\
        & + (b_{0}+b_{1})\frac{C}{\lambda_{2}} + (b_{0}+b_{1})\epsilon_2
    \end{aligned}
    \right )\ket{0}  \nonumber\\
    &\hspace{2mm} + \frac{1}{2}\left(
    \begin{aligned}
        &  -(b_{0}-b_{1}) \frac{C}{\lambda_{1}} -(b_{0}-b_{1}) \epsilon_1 \\
        & +  (b_{0}+b_{1})\frac{C}{\lambda_{2}} + (b_{0}+b_{1})\epsilon_2
    \end{aligned}
    \right )\ket{1}.  
\end{align}
The above state can be represented as \(\ket{\Psi_x}=\ket{\psi_x}+\ket{\delta}\) where 
\begin{eqnarray}
    \ket{\delta} &=& \frac{1}{2}\left\{
    (b_{0}-b_{1}) \epsilon_1 + (b_{0}+b_{1})\epsilon_2
\right\}\ket{0} \nonumber \\
& + &\frac{1}{2}\left\{-(b_{0}-b_{1}) \epsilon_1 + (b_{0}+b_{1})\epsilon_2
\right\}\ket{1}.
\end{eqnarray}
The state \(\ket{\delta}\) is the difference between the state from the algorithm with and without disorder and it only depends on the errors, \(\epsilon_i\). Since the coefficients of the register qubit is modified by an additional term depending on  \(\epsilon_i\), it is evident that the actual output is   differed by an error.

Let us now define the deviation of the actual state due to small disorder present in the controlled rotation, in a general scenario.  which we call as error, denoted by (\(E_r\)). It is defined as
\begin{equation}
E_r = \sqrt{\frac{\sum_{i=1}^{N}(\bar{x}_i - x_i)^2}{|\vec{x}|^2}},
\end{equation}
where \(\bar{x}_i\) represent the solution of the HHL algorithm without disorder. The idea is to compute quenched averaged \(\langle E_r\rangle\) which is  defined in a similar fashion as in Eq. (\ref{eq:quenched}) by replacing \(\mathcal{Q}\) by \(E_r\). 
When averaging is performed with \(10^4\) realizations having \(\expval{\epsilon}=0\) and different values of disorder strength, \(\sigma_\epsilon\), we observe that \(\expval{E_r}\) decreases with the increase of the strength of the disorder, \(\sigma_\epsilon\) (as depicted in Fig. \ref{fig:ggm_error_disorder}). On the other hand, quenched averaged GGM increases with the strength which again establishes that the genuine multiparty entanglement  does not play a beneficial role in this algorithm. Interestingly,  quenched averages of coherences of the auxiliary qubit as well as bipartite entanglement of \(\mathbf{\Lambda U}\) follows the opposite patterns of \(\langle E_r \rangle\), i.e., they decrease with the increases in the strength of the disorder as the error in the solution get enhanced due to disorder. It possibly indicate that low bipartite entanglement or coherences in subsystems deteriorates the performance of the algorithm.

\begin{figure}
\includegraphics[width=\linewidth]{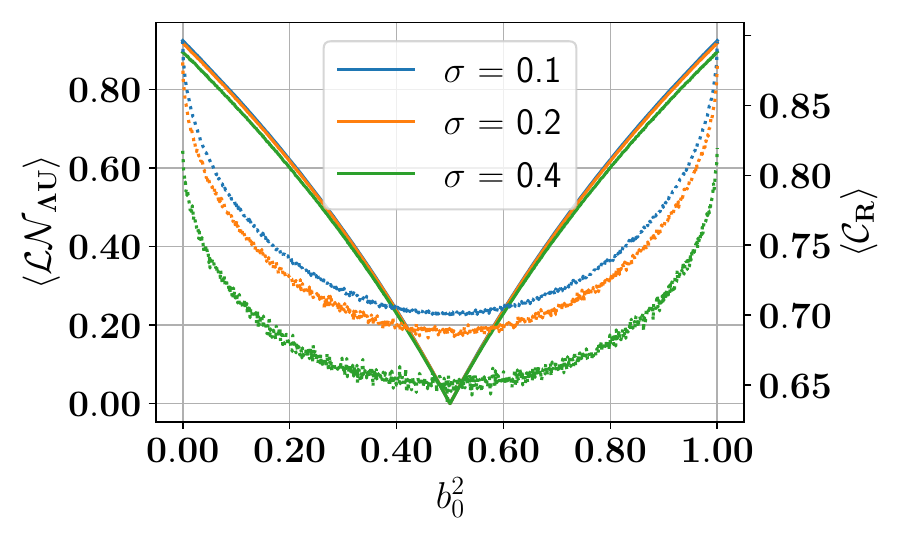}
\caption{ \textbf{Quenched average $\expval{\mathcal{C}_{\mathbf{R}}}$, and $\expval{\mathcal{LN}_{\mathbf{\Lambda U}}}$  against $b_{0}^{2}$ from Eq. (\ref{eq:2d}).}
  All other specifications are same as in Fig. \ref{fig:ggm_error_disorder}. We have seen that  the imperfections introduce errors in the output which increases proportionally with the increase of the strength of the disorder. The above figure depicts that  both $\langle \mathcal{C}_{\mathbf{R}} \rangle$ (dotted line) and $\langle \mathcal{LN}_{\mathbf{\Lambda U}} \rangle$ (solid line) decrease with the increase of the  disorder strength, which establishes that these nonclassical correlations are closely associated with the efficient functioning of the algorithm. All the axes are dimensionless.}
\label{fig:coherence_logneg_disorder}
\end{figure}

\section{Conclusion}
\label{sec:conclu}

The Harrow-Hassidim-Lloyd (HHL) algorithm is the quantum version to solve sets of linear equations and some  differential equations. In comparison to the existing classical approach, this algorithm can demonstrate exponential speed up depending on specific criteria.  As a result, determining the quantum resources responsible for speed-up is critical.

Indeed, we found that at various stages of the algorithm, quantum features of different nature,  both in composite and single systems, are produced which can be connected to the successful implementation of the algorithm. We analytically proved that for the HHL algorithm to execute successfully, the state following the controlled rotation should have nonvanishing nonclassical correlations in the form of genuine multipartite entanglement (GME) and bipartite entanglement in the reduced subsystems. Specifically, we illustrated with examples that  while GME is essential, the performance indicator and GME content behave in different ways, suggesting a negative influence for high GME on the algorithm. 
 On the other hand, bipartite entanglement between different subsystems  present in the protocol is directly proportional to the efficiency of the protocol. Moreover, we demonstrated that the quantum coherence of the readout qubit, measured by \(l_1\)-norm,  is connected with the success probability of the scheme.

We noticed that the quenched average error along with GME content in the method grows as a result of imperfect controlled rotation, which is inserted by randomly selecting parameters from a Gaussian distribution while the opposite picture emerges for quenched average bipartite entanglement  and coherence.   In the quest to establish quantum supremacy, the results obtained here have shed light on the basic ingredients to be retained in the system so that  benefits over classical algorithms can be obtained even in the presence of unavoidable defects.

\acknowledgements

We acknowledge the support from the Interdisciplinary Cyber Physical Systems (ICPS) program of the Department of Science and Technology (DST), India, Grant No.: DST/ICPS/QuST/Theme- 1/2019/23, the use of \href{https://github.com/titaschanda/QIClib}{QIClib} -- a modern C++ library for general purpose quantum information processing and quantum computing (\url{https://titaschanda.github.io/QIClib}), and the cluster computing facility at the Harish-Chandra Research Institute. 

\bibliography{bibliography}

\begin{thebibliography}{83}%
\makeatletter
\providecommand \@ifxundefined [1]{%
 \@ifx{#1\undefined}
}%
\providecommand \@ifnum [1]{%
 \ifnum #1\expandafter \@firstoftwo
 \else \expandafter \@secondoftwo
 \fi
}%
\providecommand \@ifx [1]{%
 \ifx #1\expandafter \@firstoftwo
 \else \expandafter \@secondoftwo
 \fi
}%
\providecommand \natexlab [1]{#1}%
\providecommand \enquote  [1]{``#1''}%
\providecommand \bibnamefont  [1]{#1}%
\providecommand \bibfnamefont [1]{#1}%
\providecommand \citenamefont [1]{#1}%
\providecommand \href@noop [0]{\@secondoftwo}%
\providecommand \href [0]{\begingroup \@sanitize@url \@href}%
\providecommand \@href[1]{\@@startlink{#1}\@@href}%
\providecommand \@@href[1]{\endgroup#1\@@endlink}%
\providecommand \@sanitize@url [0]{\catcode `\\12\catcode `\$12\catcode
  `\&12\catcode `\#12\catcode `\^12\catcode `\_12\catcode `\%12\relax}%
\providecommand \@@startlink[1]{}%
\providecommand \@@endlink[0]{}%
\providecommand \url  [0]{\begingroup\@sanitize@url \@url }%
\providecommand \@url [1]{\endgroup\@href {#1}{\urlprefix }}%
\providecommand \urlprefix  [0]{URL }%
\providecommand \Eprint [0]{\href }%
\providecommand \doibase [0]{http://dx.doi.org/}%
\providecommand \selectlanguage [0]{\@gobble}%
\providecommand \bibinfo  [0]{\@secondoftwo}%
\providecommand \bibfield  [0]{\@secondoftwo}%
\providecommand \translation [1]{[#1]}%
\providecommand \BibitemOpen [0]{}%
\providecommand \bibitemStop [0]{}%
\providecommand \bibitemNoStop [0]{.\EOS\space}%
\providecommand \EOS [0]{\spacefactor3000\relax}%
\providecommand \BibitemShut  [1]{\csname bibitem#1\endcsname}%
\let\auto@bib@innerbib\@empty
\bibitem [{\citenamefont {Nielsen}\ and\ \citenamefont
  {Chuang}(2010)}]{nielsen_chuang_2010}%
  \BibitemOpen
  \bibfield  {author} {\bibinfo {author} {\bibfnamefont {M.~A.}\ \bibnamefont
  {Nielsen}}\ and\ \bibinfo {author} {\bibfnamefont {I.~L.}\ \bibnamefont
  {Chuang}},\ }\href {\doibase 10.1017/CBO9780511976667} {\emph {\bibinfo
  {title} {Quantum Computation and Quantum Information: 10th Anniversary
  Edition}}}\ (\bibinfo  {publisher} {Cambridge University Press},\ \bibinfo
  {year} {2010})\BibitemShut {NoStop}%
\bibitem [{\citenamefont {Gisin}\ \emph {et~al.}(2002)\citenamefont {Gisin},
  \citenamefont {Ribordy}, \citenamefont {Tittel},\ and\ \citenamefont
  {Zbinden}}]{quantum_crypto_gisin_rmp_2002}%
  \BibitemOpen
  \bibfield  {author} {\bibinfo {author} {\bibfnamefont {N.}~\bibnamefont
  {Gisin}}, \bibinfo {author} {\bibfnamefont {G.}~\bibnamefont {Ribordy}},
  \bibinfo {author} {\bibfnamefont {W.}~\bibnamefont {Tittel}}, \ and\ \bibinfo
  {author} {\bibfnamefont {H.}~\bibnamefont {Zbinden}},\ }\href {\doibase
  10.1103/RevModPhys.74.145} {\bibfield  {journal} {\bibinfo  {journal} {Rev.
  Mod. Phys.}\ }\textbf {\bibinfo {volume} {74}},\ \bibinfo {pages} {145}
  (\bibinfo {year} {2002})}\BibitemShut {NoStop}%
\bibitem [{\citenamefont {De}\ and\ \citenamefont
  {Sen}(2011)}]{aditi_quantum_comm_review_2011}%
  \BibitemOpen
  \bibfield  {author} {\bibinfo {author} {\bibfnamefont {A.~S.}\ \bibnamefont
  {De}}\ and\ \bibinfo {author} {\bibfnamefont {U.}~\bibnamefont {Sen}},\
  }\href@noop {} {\enquote {\bibinfo {title} {Quantum advantage in
  communication networks},}\ } (\bibinfo {year} {2011}),\ \Eprint
  {http://arxiv.org/abs/1105.2412} {arXiv:1105.2412 [quant-ph]} \BibitemShut
  {NoStop}%
\bibitem [{\citenamefont {Degen}\ \emph {et~al.}(2017)\citenamefont {Degen},
  \citenamefont {Reinhard},\ and\ \citenamefont
  {Cappellaro}}]{quantum_sensing_review_2017}%
  \BibitemOpen
  \bibfield  {author} {\bibinfo {author} {\bibfnamefont {C.~L.}\ \bibnamefont
  {Degen}}, \bibinfo {author} {\bibfnamefont {F.}~\bibnamefont {Reinhard}}, \
  and\ \bibinfo {author} {\bibfnamefont {P.}~\bibnamefont {Cappellaro}},\
  }\href {\doibase 10.1103/RevModPhys.89.035002} {\bibfield  {journal}
  {\bibinfo  {journal} {Rev. Mod. Phys.}\ }\textbf {\bibinfo {volume} {89}},\
  \bibinfo {pages} {035002} (\bibinfo {year} {2017})}\BibitemShut {NoStop}%
\bibitem [{\citenamefont {Horodecki}\ \emph {et~al.}(2009)\citenamefont
  {Horodecki}, \citenamefont {Horodecki}, \citenamefont {Horodecki},\ and\
  \citenamefont {Horodecki}}]{horodecki2009}%
  \BibitemOpen
  \bibfield  {author} {\bibinfo {author} {\bibfnamefont {R.}~\bibnamefont
  {Horodecki}}, \bibinfo {author} {\bibfnamefont {P.}~\bibnamefont
  {Horodecki}}, \bibinfo {author} {\bibfnamefont {M.}~\bibnamefont
  {Horodecki}}, \ and\ \bibinfo {author} {\bibfnamefont {K.}~\bibnamefont
  {Horodecki}},\ }\href {\doibase 10.1103/RevModPhys.81.865} {\bibfield
  {journal} {\bibinfo  {journal} {Rev. Mod. Phys.}\ }\textbf {\bibinfo {volume}
  {81}},\ \bibinfo {pages} {865} (\bibinfo {year} {2009})}\BibitemShut
  {NoStop}%
\bibitem [{\citenamefont {Streltsov}\ \emph {et~al.}(2017)\citenamefont
  {Streltsov}, \citenamefont {Adesso},\ and\ \citenamefont
  {Plenio}}]{streltsov_rmp_2017}%
  \BibitemOpen
  \bibfield  {author} {\bibinfo {author} {\bibfnamefont {A.}~\bibnamefont
  {Streltsov}}, \bibinfo {author} {\bibfnamefont {G.}~\bibnamefont {Adesso}}, \
  and\ \bibinfo {author} {\bibfnamefont {M.~B.}\ \bibnamefont {Plenio}},\
  }\href {\doibase 10.1103/RevModPhys.89.041003} {\bibfield  {journal}
  {\bibinfo  {journal} {Rev. Mod. Phys.}\ }\textbf {\bibinfo {volume} {89}},\
  \bibinfo {pages} {041003} (\bibinfo {year} {2017})}\BibitemShut {NoStop}%
\bibitem [{\citenamefont {Kendon}\ and\ \citenamefont
  {Munro}(2006{\natexlab{a}})}]{kendon_qic_2006}%
  \BibitemOpen
  \bibfield  {author} {\bibinfo {author} {\bibfnamefont {V.~M.}\ \bibnamefont
  {Kendon}}\ and\ \bibinfo {author} {\bibfnamefont {W.~J.}\ \bibnamefont
  {Munro}},\ }\href@noop {} {\bibfield  {journal} {\bibinfo  {journal} {Quantum
  Info. Comput.}\ }\textbf {\bibinfo {volume} {6}},\ \bibinfo {pages}
  {630–640} (\bibinfo {year} {2006}{\natexlab{a}})}\BibitemShut {NoStop}%
\bibitem [{\citenamefont {Liu}\ \emph {et~al.}(2019)\citenamefont {Liu},
  \citenamefont {Shang},\ and\ \citenamefont {Zhang}}]{coherence_algo_2019}%
  \BibitemOpen
  \bibfield  {author} {\bibinfo {author} {\bibfnamefont {Y.-C.}\ \bibnamefont
  {Liu}}, \bibinfo {author} {\bibfnamefont {J.}~\bibnamefont {Shang}}, \ and\
  \bibinfo {author} {\bibfnamefont {X.}~\bibnamefont {Zhang}},\ }\href
  {\doibase 10.3390/e21030260} {\bibfield  {journal} {\bibinfo  {journal}
  {Entropy}\ }\textbf {\bibinfo {volume} {21}},\ \bibinfo {pages} {260}
  (\bibinfo {year} {2019})}\BibitemShut {NoStop}%
\bibitem [{\citenamefont {Grover}(1996)}]{grover_arxiv_1996}%
  \BibitemOpen
  \bibfield  {author} {\bibinfo {author} {\bibfnamefont {L.~K.}\ \bibnamefont
  {Grover}},\ }\href@noop {} {\enquote {\bibinfo {title} {A fast quantum
  mechanical algorithm for database search},}\ } (\bibinfo {year} {1996}),\
  \Eprint {http://arxiv.org/abs/quant-ph/9605043} {arXiv:quant-ph/9605043
  [quant-ph]} \BibitemShut {NoStop}%
\bibitem [{\citenamefont {Grover}(1997)}]{grover_prl_1997}%
  \BibitemOpen
  \bibfield  {author} {\bibinfo {author} {\bibfnamefont {L.~K.}\ \bibnamefont
  {Grover}},\ }\href {\doibase 10.1103/PhysRevLett.79.325} {\bibfield
  {journal} {\bibinfo  {journal} {Phys. Rev. Lett.}\ }\textbf {\bibinfo
  {volume} {79}},\ \bibinfo {pages} {325} (\bibinfo {year} {1997})}\BibitemShut
  {NoStop}%
\bibitem [{\citenamefont {Rossi}\ \emph {et~al.}(2013)\citenamefont {Rossi},
  \citenamefont {Bru\ss{}},\ and\ \citenamefont
  {Macchiavello}}]{grover_entanglement_PRA_2013}%
  \BibitemOpen
  \bibfield  {author} {\bibinfo {author} {\bibfnamefont {M.}~\bibnamefont
  {Rossi}}, \bibinfo {author} {\bibfnamefont {D.}~\bibnamefont {Bru\ss{}}}, \
  and\ \bibinfo {author} {\bibfnamefont {C.}~\bibnamefont {Macchiavello}},\
  }\href {\doibase 10.1103/PhysRevA.87.022331} {\bibfield  {journal} {\bibinfo
  {journal} {Phys. Rev. A}\ }\textbf {\bibinfo {volume} {87}},\ \bibinfo
  {pages} {022331} (\bibinfo {year} {2013})}\BibitemShut {NoStop}%
\bibitem [{\citenamefont {Pan}\ \emph {et~al.}(2017{\natexlab{a}})\citenamefont
  {Pan}, \citenamefont {Qiu},\ and\ \citenamefont {Zheng}}]{pan_qic_2017}%
  \BibitemOpen
  \bibfield  {author} {\bibinfo {author} {\bibfnamefont {M.}~\bibnamefont
  {Pan}}, \bibinfo {author} {\bibfnamefont {D.}~\bibnamefont {Qiu}}, \ and\
  \bibinfo {author} {\bibfnamefont {S.}~\bibnamefont {Zheng}},\ }\href
  {\doibase 10.1007/s11128-017-1661-4} {\bibfield  {journal} {\bibinfo
  {journal} {Quantum Inf. Process.}\ }\textbf {\bibinfo {volume} {16}},\
  \bibinfo {pages} {211} (\bibinfo {year} {2017}{\natexlab{a}})}\BibitemShut
  {NoStop}%
\bibitem [{\citenamefont {Deutsch}\ and\ \citenamefont
  {Jozsa}(1992)}]{Deutsch_Jozsa}%
  \BibitemOpen
  \bibfield  {author} {\bibinfo {author} {\bibfnamefont {D.}~\bibnamefont
  {Deutsch}}\ and\ \bibinfo {author} {\bibfnamefont {R.}~\bibnamefont
  {Jozsa}},\ }\href {\doibase 10.1098/rspa.1992.0167} {\bibfield  {journal}
  {\bibinfo  {journal} {Proc. R. Soc. Lond. A}\ }\textbf {\bibinfo {volume}
  {439}},\ \bibinfo {pages} {553} (\bibinfo {year} {1992})}\BibitemShut
  {NoStop}%
\bibitem [{\citenamefont {Cleve}\ \emph {et~al.}(1998)\citenamefont {Cleve},
  \citenamefont {Ekert}, \citenamefont {Macchiavello},\ and\ \citenamefont
  {Mosca}}]{cleve_1998}%
  \BibitemOpen
  \bibfield  {author} {\bibinfo {author} {\bibfnamefont {R.}~\bibnamefont
  {Cleve}}, \bibinfo {author} {\bibfnamefont {A.}~\bibnamefont {Ekert}},
  \bibinfo {author} {\bibfnamefont {C.}~\bibnamefont {Macchiavello}}, \ and\
  \bibinfo {author} {\bibfnamefont {M.}~\bibnamefont {Mosca}},\ }\href
  {\doibase 10.1098/rspa.1998.0164} {\bibfield  {journal} {\bibinfo  {journal}
  {Proc. R. Soc. Lond. A}\ }\textbf {\bibinfo {volume} {454}},\ \bibinfo
  {pages} {339–354} (\bibinfo {year} {1998})}\BibitemShut {NoStop}%
\bibitem [{\citenamefont {Gangopadhyay}\ \emph {et~al.}(2018)\citenamefont
  {Gangopadhyay}, \citenamefont {{Manabputra}}, \citenamefont {Behera},\ and\
  \citenamefont {Panigrahi}}]{gangopadhyay_qip_2018}%
  \BibitemOpen
  \bibfield  {author} {\bibinfo {author} {\bibfnamefont {S.}~\bibnamefont
  {Gangopadhyay}}, \bibinfo {author} {\bibnamefont {{Manabputra}}}, \bibinfo
  {author} {\bibfnamefont {B.~K.}\ \bibnamefont {Behera}}, \ and\ \bibinfo
  {author} {\bibfnamefont {P.~K.}\ \bibnamefont {Panigrahi}},\ }\href {\doibase
  10.1007/s11128-018-1932-8} {\bibfield  {journal} {\bibinfo  {journal}
  {Quantum Information Processing}\ }\textbf {\bibinfo {volume} {17}},\
  \bibinfo {pages} {160} (\bibinfo {year} {2018})}\BibitemShut {NoStop}%
\bibitem [{\citenamefont {Shor}(1994)}]{shor_1994}%
  \BibitemOpen
  \bibfield  {author} {\bibinfo {author} {\bibfnamefont {P.}~\bibnamefont
  {Shor}},\ }in\ \href {\doibase 10.1109/SFCS.1994.365700} {\emph {\bibinfo
  {booktitle} {Proceedings 35th Annual Symposium on Foundations of Computer
  Science}}}\ (\bibinfo {year} {1994})\ pp.\ \bibinfo {pages}
  {124--134}\BibitemShut {NoStop}%
\bibitem [{\citenamefont {Lu}\ \emph {et~al.}(2007)\citenamefont {Lu},
  \citenamefont {Browne}, \citenamefont {Yang},\ and\ \citenamefont
  {Pan}}]{lu_prl_2007}%
  \BibitemOpen
  \bibfield  {author} {\bibinfo {author} {\bibfnamefont {C.-Y.}\ \bibnamefont
  {Lu}}, \bibinfo {author} {\bibfnamefont {D.~E.}\ \bibnamefont {Browne}},
  \bibinfo {author} {\bibfnamefont {T.}~\bibnamefont {Yang}}, \ and\ \bibinfo
  {author} {\bibfnamefont {J.-W.}\ \bibnamefont {Pan}},\ }\href {\doibase
  10.1103/PhysRevLett.99.250504} {\bibfield  {journal} {\bibinfo  {journal}
  {Phys. Rev. Lett.}\ }\textbf {\bibinfo {volume} {99}},\ \bibinfo {pages}
  {250504} (\bibinfo {year} {2007})}\BibitemShut {NoStop}%
\bibitem [{\citenamefont {Azuma}(2018)}]{azuma_jmo_2018}%
  \BibitemOpen
  \bibfield  {author} {\bibinfo {author} {\bibfnamefont {H.}~\bibnamefont
  {Azuma}},\ }\href {\doibase 10.1080/09500340.2017.1397221} {\bibfield
  {journal} {\bibinfo  {journal} {Journal of Modern Optics}\ }\textbf {\bibinfo
  {volume} {65}},\ \bibinfo {pages} {415} (\bibinfo {year} {2018})},\ \Eprint
  {http://arxiv.org/abs/https://doi.org/10.1080/09500340.2017.1397221}
  {https://doi.org/10.1080/09500340.2017.1397221} \BibitemShut {NoStop}%
\bibitem [{\citenamefont {Naseri}\ \emph {et~al.}(2022)\citenamefont {Naseri},
  \citenamefont {Kondra}, \citenamefont {Goswami}, \citenamefont
  {Fellous-Asiani},\ and\ \citenamefont
  {Streltsov}}]{entanglement_bernstein_pra_2022}%
  \BibitemOpen
  \bibfield  {author} {\bibinfo {author} {\bibfnamefont {M.}~\bibnamefont
  {Naseri}}, \bibinfo {author} {\bibfnamefont {T.~V.}\ \bibnamefont {Kondra}},
  \bibinfo {author} {\bibfnamefont {S.}~\bibnamefont {Goswami}}, \bibinfo
  {author} {\bibfnamefont {M.}~\bibnamefont {Fellous-Asiani}}, \ and\ \bibinfo
  {author} {\bibfnamefont {A.}~\bibnamefont {Streltsov}},\ }\href {\doibase
  10.1103/PhysRevA.106.062429} {\bibfield  {journal} {\bibinfo  {journal}
  {Phys. Rev. A}\ }\textbf {\bibinfo {volume} {106}},\ \bibinfo {pages}
  {062429} (\bibinfo {year} {2022})}\BibitemShut {NoStop}%
\bibitem [{\citenamefont {Shor}(1995)}]{shor_pra_1995}%
  \BibitemOpen
  \bibfield  {author} {\bibinfo {author} {\bibfnamefont {P.~W.}\ \bibnamefont
  {Shor}},\ }\href {\doibase 10.1103/PhysRevA.52.R2493} {\bibfield  {journal}
  {\bibinfo  {journal} {Phys. Rev. A}\ }\textbf {\bibinfo {volume} {52}},\
  \bibinfo {pages} {R2493} (\bibinfo {year} {1995})}\BibitemShut {NoStop}%
\bibitem [{\citenamefont {Bru\ss{}}\ and\ \citenamefont
  {Macchiavello}(2011)}]{bruss_pra_2011}%
  \BibitemOpen
  \bibfield  {author} {\bibinfo {author} {\bibfnamefont {D.}~\bibnamefont
  {Bru\ss{}}}\ and\ \bibinfo {author} {\bibfnamefont {C.}~\bibnamefont
  {Macchiavello}},\ }\href {\doibase 10.1103/PhysRevA.83.052313} {\bibfield
  {journal} {\bibinfo  {journal} {Phys. Rev. A}\ }\textbf {\bibinfo {volume}
  {83}},\ \bibinfo {pages} {052313} (\bibinfo {year} {2011})}\BibitemShut
  {NoStop}%
\bibitem [{\citenamefont {Raussendorf}\ and\ \citenamefont
  {Briegel}(2001)}]{raussendorf_prl_2001}%
  \BibitemOpen
  \bibfield  {author} {\bibinfo {author} {\bibfnamefont {R.}~\bibnamefont
  {Raussendorf}}\ and\ \bibinfo {author} {\bibfnamefont {H.~J.}\ \bibnamefont
  {Briegel}},\ }\href {\doibase 10.1103/PhysRevLett.86.5188} {\bibfield
  {journal} {\bibinfo  {journal} {Phys. Rev. Lett.}\ }\textbf {\bibinfo
  {volume} {86}},\ \bibinfo {pages} {5188} (\bibinfo {year}
  {2001})}\BibitemShut {NoStop}%
\bibitem [{\citenamefont {Ermakov}\ and\ \citenamefont
  {Fung}(2002)}]{ermakov_pra_2002}%
  \BibitemOpen
  \bibfield  {author} {\bibinfo {author} {\bibfnamefont {V.~L.}\ \bibnamefont
  {Ermakov}}\ and\ \bibinfo {author} {\bibfnamefont {B.~M.}\ \bibnamefont
  {Fung}},\ }\href {\doibase 10.1103/PhysRevA.66.042310} {\bibfield  {journal}
  {\bibinfo  {journal} {Phys. Rev. A}\ }\textbf {\bibinfo {volume} {66}},\
  \bibinfo {pages} {042310} (\bibinfo {year} {2002})}\BibitemShut {NoStop}%
\bibitem [{\citenamefont {Xin}\ \emph {et~al.}(2018)\citenamefont {Xin},
  \citenamefont {Wang}, \citenamefont {Li}, \citenamefont {Kong}, \citenamefont
  {Wei}, \citenamefont {Wang}, \citenamefont {Ruan},\ and\ \citenamefont
  {Long}}]{Xin_cps_2018}%
  \BibitemOpen
  \bibfield  {author} {\bibinfo {author} {\bibfnamefont {T.}~\bibnamefont
  {Xin}}, \bibinfo {author} {\bibfnamefont {B.-X.}\ \bibnamefont {Wang}},
  \bibinfo {author} {\bibfnamefont {K.-R.}\ \bibnamefont {Li}}, \bibinfo
  {author} {\bibfnamefont {X.-Y.}\ \bibnamefont {Kong}}, \bibinfo {author}
  {\bibfnamefont {S.-J.}\ \bibnamefont {Wei}}, \bibinfo {author} {\bibfnamefont
  {T.}~\bibnamefont {Wang}}, \bibinfo {author} {\bibfnamefont {D.}~\bibnamefont
  {Ruan}}, \ and\ \bibinfo {author} {\bibfnamefont {G.-L.}\ \bibnamefont
  {Long}},\ }\href {\doibase 10.1088/1674-1056/27/2/020308} {\bibfield
  {journal} {\bibinfo  {journal} {Chinese Physics B}\ }\textbf {\bibinfo
  {volume} {27}},\ \bibinfo {pages} {020308} (\bibinfo {year}
  {2018})}\BibitemShut {NoStop}%
\bibitem [{\citenamefont {Xin}\ \emph {et~al.}(2020)\citenamefont {Xin},
  \citenamefont {Wei}, \citenamefont {Cui}, \citenamefont {Xiao}, \citenamefont
  {Arrazola}, \citenamefont {Lamata}, \citenamefont {Kong}, \citenamefont {Lu},
  \citenamefont {Solano},\ and\ \citenamefont {Long}}]{xin_pra_2020}%
  \BibitemOpen
  \bibfield  {author} {\bibinfo {author} {\bibfnamefont {T.}~\bibnamefont
  {Xin}}, \bibinfo {author} {\bibfnamefont {S.}~\bibnamefont {Wei}}, \bibinfo
  {author} {\bibfnamefont {J.}~\bibnamefont {Cui}}, \bibinfo {author}
  {\bibfnamefont {J.}~\bibnamefont {Xiao}}, \bibinfo {author} {\bibfnamefont
  {I.~n.}\ \bibnamefont {Arrazola}}, \bibinfo {author} {\bibfnamefont
  {L.}~\bibnamefont {Lamata}}, \bibinfo {author} {\bibfnamefont
  {X.}~\bibnamefont {Kong}}, \bibinfo {author} {\bibfnamefont {D.}~\bibnamefont
  {Lu}}, \bibinfo {author} {\bibfnamefont {E.}~\bibnamefont {Solano}}, \ and\
  \bibinfo {author} {\bibfnamefont {G.}~\bibnamefont {Long}},\ }\href {\doibase
  10.1103/PhysRevA.101.032307} {\bibfield  {journal} {\bibinfo  {journal}
  {Phys. Rev. A}\ }\textbf {\bibinfo {volume} {101}},\ \bibinfo {pages}
  {032307} (\bibinfo {year} {2020})}\BibitemShut {NoStop}%
\bibitem [{\citenamefont {Xin}\ \emph {et~al.}(2021)\citenamefont {Xin},
  \citenamefont {Che}, \citenamefont {Xi}, \citenamefont {Singh}, \citenamefont
  {Nie}, \citenamefont {Li}, \citenamefont {Dong},\ and\ \citenamefont
  {Lu}}]{xin_prl_2021}%
  \BibitemOpen
  \bibfield  {author} {\bibinfo {author} {\bibfnamefont {T.}~\bibnamefont
  {Xin}}, \bibinfo {author} {\bibfnamefont {L.}~\bibnamefont {Che}}, \bibinfo
  {author} {\bibfnamefont {C.}~\bibnamefont {Xi}}, \bibinfo {author}
  {\bibfnamefont {A.}~\bibnamefont {Singh}}, \bibinfo {author} {\bibfnamefont
  {X.}~\bibnamefont {Nie}}, \bibinfo {author} {\bibfnamefont {J.}~\bibnamefont
  {Li}}, \bibinfo {author} {\bibfnamefont {Y.}~\bibnamefont {Dong}}, \ and\
  \bibinfo {author} {\bibfnamefont {D.}~\bibnamefont {Lu}},\ }\href {\doibase
  10.1103/PhysRevLett.126.110502} {\bibfield  {journal} {\bibinfo  {journal}
  {Phys. Rev. Lett.}\ }\textbf {\bibinfo {volume} {126}},\ \bibinfo {pages}
  {110502} (\bibinfo {year} {2021})}\BibitemShut {NoStop}%
\bibitem [{\citenamefont {Gulde}\ \emph {et~al.}(2003)\citenamefont {Gulde},
  \citenamefont {Riebe}, \citenamefont {Lancaster}, \citenamefont {Becher},
  \citenamefont {Eschner}, \citenamefont {H{\"a}ffner}, \citenamefont
  {Schmidt-Kaler}, \citenamefont {Chuang},\ and\ \citenamefont
  {Blatt}}]{Gulde_nature_2003}%
  \BibitemOpen
  \bibfield  {author} {\bibinfo {author} {\bibfnamefont {S.}~\bibnamefont
  {Gulde}}, \bibinfo {author} {\bibfnamefont {M.}~\bibnamefont {Riebe}},
  \bibinfo {author} {\bibfnamefont {G.~P.~T.}\ \bibnamefont {Lancaster}},
  \bibinfo {author} {\bibfnamefont {C.}~\bibnamefont {Becher}}, \bibinfo
  {author} {\bibfnamefont {J.}~\bibnamefont {Eschner}}, \bibinfo {author}
  {\bibfnamefont {H.}~\bibnamefont {H{\"a}ffner}}, \bibinfo {author}
  {\bibfnamefont {F.}~\bibnamefont {Schmidt-Kaler}}, \bibinfo {author}
  {\bibfnamefont {I.~L.}\ \bibnamefont {Chuang}}, \ and\ \bibinfo {author}
  {\bibfnamefont {R.}~\bibnamefont {Blatt}},\ }\href {\doibase
  10.1038/nature01336} {\bibfield  {journal} {\bibinfo  {journal} {Nature}\
  }\textbf {\bibinfo {volume} {421}},\ \bibinfo {pages} {48} (\bibinfo {year}
  {2003})}\BibitemShut {NoStop}%
\bibitem [{\citenamefont {Graham}\ \emph {et~al.}(2022)\citenamefont {Graham},
  \citenamefont {Song}, \citenamefont {Scott}, \citenamefont {Poole},
  \citenamefont {Phuttitarn}, \citenamefont {Jooya}, \citenamefont {Eichler},
  \citenamefont {Jiang}, \citenamefont {Marra}, \citenamefont {Grinkemeyer},
  \citenamefont {Kwon}, \citenamefont {Ebert}, \citenamefont {Cherek},
  \citenamefont {Lichtman}, \citenamefont {Gillette}, \citenamefont {Gilbert},
  \citenamefont {Bowman}, \citenamefont {Ballance}, \citenamefont {Campbell},
  \citenamefont {Dahl}, \citenamefont {Crawford}, \citenamefont {Blunt},
  \citenamefont {Rogers}, \citenamefont {Noel},\ and\ \citenamefont
  {Saffman}}]{neutralatom_phaseestimation_2022}%
  \BibitemOpen
  \bibfield  {author} {\bibinfo {author} {\bibfnamefont {T.~M.}\ \bibnamefont
  {Graham}}, \bibinfo {author} {\bibfnamefont {Y.}~\bibnamefont {Song}},
  \bibinfo {author} {\bibfnamefont {J.}~\bibnamefont {Scott}}, \bibinfo
  {author} {\bibfnamefont {C.}~\bibnamefont {Poole}}, \bibinfo {author}
  {\bibfnamefont {L.}~\bibnamefont {Phuttitarn}}, \bibinfo {author}
  {\bibfnamefont {K.}~\bibnamefont {Jooya}}, \bibinfo {author} {\bibfnamefont
  {P.}~\bibnamefont {Eichler}}, \bibinfo {author} {\bibfnamefont
  {X.}~\bibnamefont {Jiang}}, \bibinfo {author} {\bibfnamefont
  {A.}~\bibnamefont {Marra}}, \bibinfo {author} {\bibfnamefont
  {B.}~\bibnamefont {Grinkemeyer}}, \bibinfo {author} {\bibfnamefont
  {M.}~\bibnamefont {Kwon}}, \bibinfo {author} {\bibfnamefont {M.}~\bibnamefont
  {Ebert}}, \bibinfo {author} {\bibfnamefont {J.}~\bibnamefont {Cherek}},
  \bibinfo {author} {\bibfnamefont {M.~T.}\ \bibnamefont {Lichtman}}, \bibinfo
  {author} {\bibfnamefont {M.}~\bibnamefont {Gillette}}, \bibinfo {author}
  {\bibfnamefont {J.}~\bibnamefont {Gilbert}}, \bibinfo {author} {\bibfnamefont
  {D.}~\bibnamefont {Bowman}}, \bibinfo {author} {\bibfnamefont
  {T.}~\bibnamefont {Ballance}}, \bibinfo {author} {\bibfnamefont
  {C.}~\bibnamefont {Campbell}}, \bibinfo {author} {\bibfnamefont {E.~D.}\
  \bibnamefont {Dahl}}, \bibinfo {author} {\bibfnamefont {O.}~\bibnamefont
  {Crawford}}, \bibinfo {author} {\bibfnamefont {N.~S.}\ \bibnamefont {Blunt}},
  \bibinfo {author} {\bibfnamefont {B.}~\bibnamefont {Rogers}}, \bibinfo
  {author} {\bibfnamefont {T.}~\bibnamefont {Noel}}, \ and\ \bibinfo {author}
  {\bibfnamefont {M.}~\bibnamefont {Saffman}},\ }\href {\doibase
  10.1038/s41586-022-04603-6} {\bibfield  {journal} {\bibinfo  {journal}
  {Nature}\ }\textbf {\bibinfo {volume} {604}},\ \bibinfo {pages} {457}
  (\bibinfo {year} {2022})}\BibitemShut {NoStop}%
\bibitem [{\citenamefont {Zheng}\ \emph {et~al.}(2017)\citenamefont {Zheng},
  \citenamefont {Song}, \citenamefont {Chen}, \citenamefont {Xia},
  \citenamefont {Liu}, \citenamefont {Guo}, \citenamefont {Zhang},
  \citenamefont {Xu}, \citenamefont {Deng}, \citenamefont {Huang},
  \citenamefont {Wu}, \citenamefont {Yan}, \citenamefont {Zheng}, \citenamefont
  {Lu}, \citenamefont {Pan}, \citenamefont {Wang}, \citenamefont {Lu},\ and\
  \citenamefont {Zhu}}]{zheng_prl_2017}%
  \BibitemOpen
  \bibfield  {author} {\bibinfo {author} {\bibfnamefont {Y.}~\bibnamefont
  {Zheng}}, \bibinfo {author} {\bibfnamefont {C.}~\bibnamefont {Song}},
  \bibinfo {author} {\bibfnamefont {M.-C.}\ \bibnamefont {Chen}}, \bibinfo
  {author} {\bibfnamefont {B.}~\bibnamefont {Xia}}, \bibinfo {author}
  {\bibfnamefont {W.}~\bibnamefont {Liu}}, \bibinfo {author} {\bibfnamefont
  {Q.}~\bibnamefont {Guo}}, \bibinfo {author} {\bibfnamefont {L.}~\bibnamefont
  {Zhang}}, \bibinfo {author} {\bibfnamefont {D.}~\bibnamefont {Xu}}, \bibinfo
  {author} {\bibfnamefont {H.}~\bibnamefont {Deng}}, \bibinfo {author}
  {\bibfnamefont {K.}~\bibnamefont {Huang}}, \bibinfo {author} {\bibfnamefont
  {Y.}~\bibnamefont {Wu}}, \bibinfo {author} {\bibfnamefont {Z.}~\bibnamefont
  {Yan}}, \bibinfo {author} {\bibfnamefont {D.}~\bibnamefont {Zheng}}, \bibinfo
  {author} {\bibfnamefont {L.}~\bibnamefont {Lu}}, \bibinfo {author}
  {\bibfnamefont {J.-W.}\ \bibnamefont {Pan}}, \bibinfo {author} {\bibfnamefont
  {H.}~\bibnamefont {Wang}}, \bibinfo {author} {\bibfnamefont {C.-Y.}\
  \bibnamefont {Lu}}, \ and\ \bibinfo {author} {\bibfnamefont {X.}~\bibnamefont
  {Zhu}},\ }\href {\doibase 10.1103/PhysRevLett.118.210504} {\bibfield
  {journal} {\bibinfo  {journal} {Phys. Rev. Lett.}\ }\textbf {\bibinfo
  {volume} {118}},\ \bibinfo {pages} {210504} (\bibinfo {year}
  {2017})}\BibitemShut {NoStop}%
\bibitem [{\citenamefont {C\'orcoles}\ \emph {et~al.}(2021)\citenamefont
  {C\'orcoles}, \citenamefont {Takita}, \citenamefont {Inoue}, \citenamefont
  {Lekuch}, \citenamefont {Minev}, \citenamefont {Chow},\ and\ \citenamefont
  {Gambetta}}]{takita_prl_2021}%
  \BibitemOpen
  \bibfield  {author} {\bibinfo {author} {\bibfnamefont {A.~D.}\ \bibnamefont
  {C\'orcoles}}, \bibinfo {author} {\bibfnamefont {M.}~\bibnamefont {Takita}},
  \bibinfo {author} {\bibfnamefont {K.}~\bibnamefont {Inoue}}, \bibinfo
  {author} {\bibfnamefont {S.}~\bibnamefont {Lekuch}}, \bibinfo {author}
  {\bibfnamefont {Z.~K.}\ \bibnamefont {Minev}}, \bibinfo {author}
  {\bibfnamefont {J.~M.}\ \bibnamefont {Chow}}, \ and\ \bibinfo {author}
  {\bibfnamefont {J.~M.}\ \bibnamefont {Gambetta}},\ }\href {\doibase
  10.1103/PhysRevLett.127.100501} {\bibfield  {journal} {\bibinfo  {journal}
  {Phys. Rev. Lett.}\ }\textbf {\bibinfo {volume} {127}},\ \bibinfo {pages}
  {100501} (\bibinfo {year} {2021})}\BibitemShut {NoStop}%
\bibitem [{\citenamefont {Lanyon}\ \emph
  {et~al.}(2007{\natexlab{a}})\citenamefont {Lanyon}, \citenamefont {Weinhold},
  \citenamefont {Langford}, \citenamefont {Barbieri}, \citenamefont {James},
  \citenamefont {Gilchrist},\ and\ \citenamefont
  {White}}]{shor_experi_photonics_2007}%
  \BibitemOpen
  \bibfield  {author} {\bibinfo {author} {\bibfnamefont {B.~P.}\ \bibnamefont
  {Lanyon}}, \bibinfo {author} {\bibfnamefont {T.~J.}\ \bibnamefont
  {Weinhold}}, \bibinfo {author} {\bibfnamefont {N.~K.}\ \bibnamefont
  {Langford}}, \bibinfo {author} {\bibfnamefont {M.}~\bibnamefont {Barbieri}},
  \bibinfo {author} {\bibfnamefont {D.~F.~V.}\ \bibnamefont {James}}, \bibinfo
  {author} {\bibfnamefont {A.}~\bibnamefont {Gilchrist}}, \ and\ \bibinfo
  {author} {\bibfnamefont {A.~G.}\ \bibnamefont {White}},\ }\href {\doibase
  10.1103/PhysRevLett.99.250505} {\bibfield  {journal} {\bibinfo  {journal}
  {Phys. Rev. Lett.}\ }\textbf {\bibinfo {volume} {99}},\ \bibinfo {pages}
  {250505} (\bibinfo {year} {2007}{\natexlab{a}})}\BibitemShut {NoStop}%
\bibitem [{\citenamefont {Watson}\ \emph {et~al.}(2018)\citenamefont {Watson},
  \citenamefont {Philips}, \citenamefont {Kawakami}, \citenamefont {Ward},
  \citenamefont {Scarlino}, \citenamefont {Veldhorst}, \citenamefont {Savage},
  \citenamefont {Lagally}, \citenamefont {Friesen}, \citenamefont
  {Coppersmith}, \citenamefont {Eriksson},\ and\ \citenamefont
  {Vandersypen}}]{silicon_experimental_deutsch_grover_2018}%
  \BibitemOpen
  \bibfield  {author} {\bibinfo {author} {\bibfnamefont {T.~F.}\ \bibnamefont
  {Watson}}, \bibinfo {author} {\bibfnamefont {S.~G.~J.}\ \bibnamefont
  {Philips}}, \bibinfo {author} {\bibfnamefont {E.}~\bibnamefont {Kawakami}},
  \bibinfo {author} {\bibfnamefont {D.~R.}\ \bibnamefont {Ward}}, \bibinfo
  {author} {\bibfnamefont {P.}~\bibnamefont {Scarlino}}, \bibinfo {author}
  {\bibfnamefont {M.}~\bibnamefont {Veldhorst}}, \bibinfo {author}
  {\bibfnamefont {D.~E.}\ \bibnamefont {Savage}}, \bibinfo {author}
  {\bibfnamefont {M.~G.}\ \bibnamefont {Lagally}}, \bibinfo {author}
  {\bibfnamefont {M.}~\bibnamefont {Friesen}}, \bibinfo {author} {\bibfnamefont
  {S.~N.}\ \bibnamefont {Coppersmith}}, \bibinfo {author} {\bibfnamefont
  {M.~A.}\ \bibnamefont {Eriksson}}, \ and\ \bibinfo {author} {\bibfnamefont
  {L.~M.~K.}\ \bibnamefont {Vandersypen}},\ }\href {\doibase
  10.1038/nature25766} {\bibfield  {journal} {\bibinfo  {journal} {Nature}\
  }\textbf {\bibinfo {volume} {555}},\ \bibinfo {pages} {633} (\bibinfo {year}
  {2018})}\BibitemShut {NoStop}%
\bibitem [{\citenamefont {Lanyon}\ \emph
  {et~al.}(2007{\natexlab{b}})\citenamefont {Lanyon}, \citenamefont {Weinhold},
  \citenamefont {Langford}, \citenamefont {Barbieri}, \citenamefont {James},
  \citenamefont {Gilchrist},\ and\ \citenamefont {White}}]{lanyon_prl_2007}%
  \BibitemOpen
  \bibfield  {author} {\bibinfo {author} {\bibfnamefont {B.~P.}\ \bibnamefont
  {Lanyon}}, \bibinfo {author} {\bibfnamefont {T.~J.}\ \bibnamefont
  {Weinhold}}, \bibinfo {author} {\bibfnamefont {N.~K.}\ \bibnamefont
  {Langford}}, \bibinfo {author} {\bibfnamefont {M.}~\bibnamefont {Barbieri}},
  \bibinfo {author} {\bibfnamefont {D.~F.~V.}\ \bibnamefont {James}}, \bibinfo
  {author} {\bibfnamefont {A.}~\bibnamefont {Gilchrist}}, \ and\ \bibinfo
  {author} {\bibfnamefont {A.~G.}\ \bibnamefont {White}},\ }\href {\doibase
  10.1103/PhysRevLett.99.250505} {\bibfield  {journal} {\bibinfo  {journal}
  {Phys. Rev. Lett.}\ }\textbf {\bibinfo {volume} {99}},\ \bibinfo {pages}
  {250505} (\bibinfo {year} {2007}{\natexlab{b}})}\BibitemShut {NoStop}%
\bibitem [{\citenamefont {Havl{\'i}{\v{c}}ek}\ \emph
  {et~al.}(2019)\citenamefont {Havl{\'i}{\v{c}}ek}, \citenamefont
  {C{\'o}rcoles}, \citenamefont {Temme}, \citenamefont {Harrow}, \citenamefont
  {Kandala}, \citenamefont {Chow},\ and\ \citenamefont
  {Gambetta}}]{Havlicek_natute_2019}%
  \BibitemOpen
  \bibfield  {author} {\bibinfo {author} {\bibfnamefont {V.}~\bibnamefont
  {Havl{\'i}{\v{c}}ek}}, \bibinfo {author} {\bibfnamefont {A.~D.}\ \bibnamefont
  {C{\'o}rcoles}}, \bibinfo {author} {\bibfnamefont {K.}~\bibnamefont {Temme}},
  \bibinfo {author} {\bibfnamefont {A.~W.}\ \bibnamefont {Harrow}}, \bibinfo
  {author} {\bibfnamefont {A.}~\bibnamefont {Kandala}}, \bibinfo {author}
  {\bibfnamefont {J.~M.}\ \bibnamefont {Chow}}, \ and\ \bibinfo {author}
  {\bibfnamefont {J.~M.}\ \bibnamefont {Gambetta}},\ }\href {\doibase
  10.1038/s41586-019-0980-2} {\bibfield  {journal} {\bibinfo  {journal}
  {Nature}\ }\textbf {\bibinfo {volume} {567}},\ \bibinfo {pages} {209}
  (\bibinfo {year} {2019})}\BibitemShut {NoStop}%
\bibitem [{\citenamefont {Bennett}\ and\ \citenamefont
  {Wiesner}(1992)}]{bennett_prl_1992}%
  \BibitemOpen
  \bibfield  {author} {\bibinfo {author} {\bibfnamefont {C.~H.}\ \bibnamefont
  {Bennett}}\ and\ \bibinfo {author} {\bibfnamefont {S.~J.}\ \bibnamefont
  {Wiesner}},\ }\href {\doibase 10.1103/PhysRevLett.69.2881} {\bibfield
  {journal} {\bibinfo  {journal} {Phys. Rev. Lett.}\ }\textbf {\bibinfo
  {volume} {69}},\ \bibinfo {pages} {2881} (\bibinfo {year}
  {1992})}\BibitemShut {NoStop}%
\bibitem [{\citenamefont {Bennett}\ \emph {et~al.}(1993)\citenamefont
  {Bennett}, \citenamefont {Brassard}, \citenamefont {Cr\'epeau}, \citenamefont
  {Jozsa}, \citenamefont {Peres},\ and\ \citenamefont
  {Wootters}}]{benett_prl_1993}%
  \BibitemOpen
  \bibfield  {author} {\bibinfo {author} {\bibfnamefont {C.~H.}\ \bibnamefont
  {Bennett}}, \bibinfo {author} {\bibfnamefont {G.}~\bibnamefont {Brassard}},
  \bibinfo {author} {\bibfnamefont {C.}~\bibnamefont {Cr\'epeau}}, \bibinfo
  {author} {\bibfnamefont {R.}~\bibnamefont {Jozsa}}, \bibinfo {author}
  {\bibfnamefont {A.}~\bibnamefont {Peres}}, \ and\ \bibinfo {author}
  {\bibfnamefont {W.~K.}\ \bibnamefont {Wootters}},\ }\href {\doibase
  10.1103/PhysRevLett.70.1895} {\bibfield  {journal} {\bibinfo  {journal}
  {Phys. Rev. Lett.}\ }\textbf {\bibinfo {volume} {70}},\ \bibinfo {pages}
  {1895} (\bibinfo {year} {1993})}\BibitemShut {NoStop}%
\bibitem [{\citenamefont {Ekert}(1991)}]{ekert1991}%
  \BibitemOpen
  \bibfield  {author} {\bibinfo {author} {\bibfnamefont {A.~K.}\ \bibnamefont
  {Ekert}},\ }\href {\doibase 10.1103/PhysRevLett.67.661} {\bibfield  {journal}
  {\bibinfo  {journal} {Phys. Rev. Lett.}\ }\textbf {\bibinfo {volume} {67}},\
  \bibinfo {pages} {661} (\bibinfo {year} {1991})}\BibitemShut {NoStop}%
\bibitem [{\citenamefont {Amico}\ \emph {et~al.}(2008)\citenamefont {Amico},
  \citenamefont {Fazio}, \citenamefont {Osterloh},\ and\ \citenamefont
  {Vedral}}]{amico_fazio_rmp_2008}%
  \BibitemOpen
  \bibfield  {author} {\bibinfo {author} {\bibfnamefont {L.}~\bibnamefont
  {Amico}}, \bibinfo {author} {\bibfnamefont {R.}~\bibnamefont {Fazio}},
  \bibinfo {author} {\bibfnamefont {A.}~\bibnamefont {Osterloh}}, \ and\
  \bibinfo {author} {\bibfnamefont {V.}~\bibnamefont {Vedral}},\ }\href
  {\doibase 10.1103/RevModPhys.80.517} {\bibfield  {journal} {\bibinfo
  {journal} {Rev. Mod. Phys.}\ }\textbf {\bibinfo {volume} {80}},\ \bibinfo
  {pages} {517} (\bibinfo {year} {2008})}\BibitemShut {NoStop}%
\bibitem [{\citenamefont {Lewenstein}\ \emph {et~al.}(2007)\citenamefont
  {Lewenstein}, \citenamefont {Sanpera}, \citenamefont {Ahufinger},
  \citenamefont {Damski}, \citenamefont {Sen(De)},\ and\ \citenamefont
  {Sen}}]{aditi_advanves_physics_2007}%
  \BibitemOpen
  \bibfield  {author} {\bibinfo {author} {\bibfnamefont {M.}~\bibnamefont
  {Lewenstein}}, \bibinfo {author} {\bibfnamefont {A.}~\bibnamefont {Sanpera}},
  \bibinfo {author} {\bibfnamefont {V.}~\bibnamefont {Ahufinger}}, \bibinfo
  {author} {\bibfnamefont {B.}~\bibnamefont {Damski}}, \bibinfo {author}
  {\bibfnamefont {A.}~\bibnamefont {Sen(De)}}, \ and\ \bibinfo {author}
  {\bibfnamefont {U.}~\bibnamefont {Sen}},\ }\href {\doibase
  10.1080/00018730701223200} {\bibfield  {journal} {\bibinfo  {journal} {Adv.
  Phys.}\ }\textbf {\bibinfo {volume} {56}},\ \bibinfo {pages} {243} (\bibinfo
  {year} {2007})}\BibitemShut {NoStop}%
\bibitem [{\citenamefont {Abanin}\ \emph {et~al.}(2019)\citenamefont {Abanin},
  \citenamefont {Altman}, \citenamefont {Bloch},\ and\ \citenamefont
  {Serbyn}}]{bloch_rmp_2019}%
  \BibitemOpen
  \bibfield  {author} {\bibinfo {author} {\bibfnamefont {D.~A.}\ \bibnamefont
  {Abanin}}, \bibinfo {author} {\bibfnamefont {E.}~\bibnamefont {Altman}},
  \bibinfo {author} {\bibfnamefont {I.}~\bibnamefont {Bloch}}, \ and\ \bibinfo
  {author} {\bibfnamefont {M.}~\bibnamefont {Serbyn}},\ }\href {\doibase
  10.1103/RevModPhys.91.021001} {\bibfield  {journal} {\bibinfo  {journal}
  {Rev. Mod. Phys.}\ }\textbf {\bibinfo {volume} {91}},\ \bibinfo {pages}
  {021001} (\bibinfo {year} {2019})}\BibitemShut {NoStop}%
\bibitem [{\citenamefont {Fujikawa}\ \emph {et~al.}(2019)\citenamefont
  {Fujikawa}, \citenamefont {Oh},\ and\ \citenamefont
  {Umetsu}}]{ent_cohe_grover_2019}%
  \BibitemOpen
  \bibfield  {author} {\bibinfo {author} {\bibfnamefont {K.}~\bibnamefont
  {Fujikawa}}, \bibinfo {author} {\bibfnamefont {C.~H.}\ \bibnamefont {Oh}}, \
  and\ \bibinfo {author} {\bibfnamefont {K.}~\bibnamefont {Umetsu}},\ }\href
  {\doibase 10.1142/S0217732319501463} {\bibfield  {journal} {\bibinfo
  {journal} {Mod. Phys. Lett. A}\ }\textbf {\bibinfo {volume} {34}},\ \bibinfo
  {pages} {1950146} (\bibinfo {year} {2019})}\BibitemShut {NoStop}%
\bibitem [{\citenamefont {Hillery}(2016)}]{hillery_pra_2016}%
  \BibitemOpen
  \bibfield  {author} {\bibinfo {author} {\bibfnamefont {M.}~\bibnamefont
  {Hillery}},\ }\href {\doibase 10.1103/PhysRevA.93.012111} {\bibfield
  {journal} {\bibinfo  {journal} {Phys. Rev. A}\ }\textbf {\bibinfo {volume}
  {93}},\ \bibinfo {pages} {012111} (\bibinfo {year} {2016})}\BibitemShut
  {NoStop}%
\bibitem [{\citenamefont {Pan}\ \emph {et~al.}(2017{\natexlab{b}})\citenamefont
  {Pan}, \citenamefont {Qiu},\ and\ \citenamefont
  {Zheng}}]{global_ent_grover_2017}%
  \BibitemOpen
  \bibfield  {author} {\bibinfo {author} {\bibfnamefont {M.}~\bibnamefont
  {Pan}}, \bibinfo {author} {\bibfnamefont {D.}~\bibnamefont {Qiu}}, \ and\
  \bibinfo {author} {\bibfnamefont {S.}~\bibnamefont {Zheng}},\ }\href
  {\doibase 10.1007/s11128-017-1661-4} {\bibfield  {journal} {\bibinfo
  {journal} {Quantum Inf. Process.}\ }\textbf {\bibinfo {volume} {16}},\
  \bibinfo {pages} {1} (\bibinfo {year} {2017}{\natexlab{b}})}\BibitemShut
  {NoStop}%
\bibitem [{\citenamefont {Qu}\ \emph {et~al.}(2012)\citenamefont {Qu},
  \citenamefont {Wang}, \citenamefont {shang Li}, \citenamefont {li~Zhao},
  \citenamefont {ru~Bao},\ and\ \citenamefont {chun
  Cao}}]{multi_ent_grover_2012}%
  \BibitemOpen
  \bibfield  {author} {\bibinfo {author} {\bibfnamefont {R.}~\bibnamefont
  {Qu}}, \bibinfo {author} {\bibfnamefont {J.}~\bibnamefont {Wang}}, \bibinfo
  {author} {\bibfnamefont {Z.}~\bibnamefont {shang Li}}, \bibinfo {author}
  {\bibfnamefont {S.}~\bibnamefont {li~Zhao}}, \bibinfo {author} {\bibfnamefont
  {Y.}~\bibnamefont {ru~Bao}}, \ and\ \bibinfo {author} {\bibfnamefont
  {X.}~\bibnamefont {chun Cao}},\ }\href@noop {} {\enquote {\bibinfo {title}
  {Multipartite entanglement and grover's search algorithm},}\ } (\bibinfo
  {year} {2012}),\ \Eprint {http://arxiv.org/abs/1210.3418} {arXiv:1210.3418
  [quant-ph]} \BibitemShut {NoStop}%
\bibitem [{\citenamefont {Kendon}\ and\ \citenamefont
  {Munro}(2006{\natexlab{b}})}]{shors_entanglement_2006}%
  \BibitemOpen
  \bibfield  {author} {\bibinfo {author} {\bibfnamefont {V.~M.}\ \bibnamefont
  {Kendon}}\ and\ \bibinfo {author} {\bibfnamefont {W.~J.}\ \bibnamefont
  {Munro}},\ }\href@noop {} {\bibfield  {journal} {\bibinfo  {journal} {Quantum
  Info. Comput.}\ }\textbf {\bibinfo {volume} {6}},\ \bibinfo {pages}
  {630–640} (\bibinfo {year} {2006}{\natexlab{b}})}\BibitemShut {NoStop}%
\bibitem [{\citenamefont {Harrow}\ \emph {et~al.}(2009)\citenamefont {Harrow},
  \citenamefont {Hassidim},\ and\ \citenamefont {Lloyd}}]{HHL_PRL_2009}%
  \BibitemOpen
  \bibfield  {author} {\bibinfo {author} {\bibfnamefont {A.~W.}\ \bibnamefont
  {Harrow}}, \bibinfo {author} {\bibfnamefont {A.}~\bibnamefont {Hassidim}}, \
  and\ \bibinfo {author} {\bibfnamefont {S.}~\bibnamefont {Lloyd}},\ }\href
  {\doibase 10.1103/PhysRevLett.103.150502} {\bibfield  {journal} {\bibinfo
  {journal} {Phys. Rev. Lett.}\ }\textbf {\bibinfo {volume} {103}},\ \bibinfo
  {pages} {150502} (\bibinfo {year} {2009})}\BibitemShut {NoStop}%
\bibitem [{\citenamefont {Cao}\ \emph {et~al.}(2013)\citenamefont {Cao},
  \citenamefont {Papageorgiou}, \citenamefont {Petras}, \citenamefont {Traub},\
  and\ \citenamefont {Kais}}]{cao_njp_2013}%
  \BibitemOpen
  \bibfield  {author} {\bibinfo {author} {\bibfnamefont {Y.}~\bibnamefont
  {Cao}}, \bibinfo {author} {\bibfnamefont {A.}~\bibnamefont {Papageorgiou}},
  \bibinfo {author} {\bibfnamefont {I.}~\bibnamefont {Petras}}, \bibinfo
  {author} {\bibfnamefont {J.}~\bibnamefont {Traub}}, \ and\ \bibinfo {author}
  {\bibfnamefont {S.}~\bibnamefont {Kais}},\ }\href {\doibase
  10.1088/1367-2630/15/1/013021} {\bibfield  {journal} {\bibinfo  {journal}
  {New Journal of Physics}\ }\textbf {\bibinfo {volume} {15}},\ \bibinfo
  {pages} {013021} (\bibinfo {year} {2013})}\BibitemShut {NoStop}%
\bibitem [{\citenamefont {Cao}\ \emph {et~al.}(2012)\citenamefont {Cao},
  \citenamefont {Daskin}, \citenamefont {Frankel},\ and\ \citenamefont
  {Kais}}]{Yudong_mp_2012}%
  \BibitemOpen
  \bibfield  {author} {\bibinfo {author} {\bibfnamefont {Y.}~\bibnamefont
  {Cao}}, \bibinfo {author} {\bibfnamefont {A.}~\bibnamefont {Daskin}},
  \bibinfo {author} {\bibfnamefont {S.}~\bibnamefont {Frankel}}, \ and\
  \bibinfo {author} {\bibfnamefont {S.}~\bibnamefont {Kais}},\ }\href {\doibase
  10.1080/00268976.2012.668289} {\bibfield  {journal} {\bibinfo  {journal}
  {Molecular Physics}\ }\textbf {\bibinfo {volume} {110}},\ \bibinfo {pages}
  {1675} (\bibinfo {year} {2012})},\ \Eprint
  {http://arxiv.org/abs/https://doi.org/10.1080/00268976.2012.668289}
  {https://doi.org/10.1080/00268976.2012.668289} \BibitemShut {NoStop}%
\bibitem [{\citenamefont {au2}\ \emph {et~al.}(2023)\citenamefont {au2},
  \citenamefont {Zaman},\ and\ \citenamefont {Wong}}]{morrell_arxiv_2023}%
  \BibitemOpen
  \bibfield  {author} {\bibinfo {author} {\bibfnamefont {H.~J. M.~J.}\
  \bibnamefont {au2}}, \bibinfo {author} {\bibfnamefont {A.}~\bibnamefont
  {Zaman}}, \ and\ \bibinfo {author} {\bibfnamefont {H.~Y.}\ \bibnamefont
  {Wong}},\ }\href@noop {} {\enquote {\bibinfo {title} {Step-by-step hhl
  algorithm walkthrough to enhance the understanding of critical quantum
  computing concepts},}\ } (\bibinfo {year} {2023}),\ \Eprint
  {http://arxiv.org/abs/2108.09004} {arXiv:2108.09004 [quant-ph]} \BibitemShut
  {NoStop}%
\bibitem [{\citenamefont {Hwang}\ \emph {et~al.}(2022)\citenamefont {Hwang},
  \citenamefont {Kim}, \citenamefont {Jung}, \citenamefont {Woo},\ and\
  \citenamefont {Park}}]{hwang_arxiv_2022}%
  \BibitemOpen
  \bibfield  {author} {\bibinfo {author} {\bibfnamefont {M.-R.}\ \bibnamefont
  {Hwang}}, \bibinfo {author} {\bibfnamefont {M.}~\bibnamefont {Kim}}, \bibinfo
  {author} {\bibfnamefont {E.}~\bibnamefont {Jung}}, \bibinfo {author}
  {\bibfnamefont {C.-Y.}\ \bibnamefont {Woo}}, \ and\ \bibinfo {author}
  {\bibfnamefont {D.}~\bibnamefont {Park}},\ }\href@noop {} {\enquote {\bibinfo
  {title} {Tripartite entanglement and matrix inversion quantum algorithm},}\ }
  (\bibinfo {year} {2022}),\ \Eprint {http://arxiv.org/abs/2203.10780}
  {arXiv:2203.10780 [quant-ph]} \BibitemShut {NoStop}%
\bibitem [{\citenamefont {Feng}\ \emph {et~al.}(2022)\citenamefont {Feng},
  \citenamefont {Chen},\ and\ \citenamefont {Zhao}}]{ent_cohe_HHL_2022}%
  \BibitemOpen
  \bibfield  {author} {\bibinfo {author} {\bibfnamefont {C.}~\bibnamefont
  {Feng}}, \bibinfo {author} {\bibfnamefont {L.}~\bibnamefont {Chen}}, \ and\
  \bibinfo {author} {\bibfnamefont {L.}~\bibnamefont {Zhao}},\ }\href@noop {}
  {\enquote {\bibinfo {title} {Coherence and entanglement in grover and
  harrow-hassidim-lloyd algorithm},}\ } (\bibinfo {year} {2022}),\ \Eprint
  {http://arxiv.org/abs/2212.13938} {arXiv:2212.13938 [quant-ph]} \BibitemShut
  {NoStop}%
\bibitem [{\citenamefont {Pan}\ \emph {et~al.}(2014)\citenamefont {Pan},
  \citenamefont {Cao}, \citenamefont {Yao}, \citenamefont {Li}, \citenamefont
  {Ju}, \citenamefont {Chen}, \citenamefont {Peng}, \citenamefont {Kais},\ and\
  \citenamefont {Du}}]{pan_pra_2014}%
  \BibitemOpen
  \bibfield  {author} {\bibinfo {author} {\bibfnamefont {J.}~\bibnamefont
  {Pan}}, \bibinfo {author} {\bibfnamefont {Y.}~\bibnamefont {Cao}}, \bibinfo
  {author} {\bibfnamefont {X.}~\bibnamefont {Yao}}, \bibinfo {author}
  {\bibfnamefont {Z.}~\bibnamefont {Li}}, \bibinfo {author} {\bibfnamefont
  {C.}~\bibnamefont {Ju}}, \bibinfo {author} {\bibfnamefont {H.}~\bibnamefont
  {Chen}}, \bibinfo {author} {\bibfnamefont {X.}~\bibnamefont {Peng}}, \bibinfo
  {author} {\bibfnamefont {S.}~\bibnamefont {Kais}}, \ and\ \bibinfo {author}
  {\bibfnamefont {J.}~\bibnamefont {Du}},\ }\href {\doibase
  10.1103/PhysRevA.89.022313} {\bibfield  {journal} {\bibinfo  {journal} {Phys.
  Rev. A}\ }\textbf {\bibinfo {volume} {89}},\ \bibinfo {pages} {022313}
  (\bibinfo {year} {2014})}\BibitemShut {NoStop}%
\bibitem [{\citenamefont {Cai}\ \emph {et~al.}(2013)\citenamefont {Cai},
  \citenamefont {Weedbrook}, \citenamefont {Su}, \citenamefont {Chen},
  \citenamefont {Gu}, \citenamefont {Zhu}, \citenamefont {Li}, \citenamefont
  {Liu}, \citenamefont {Lu},\ and\ \citenamefont
  {Pan}}]{pan_prl_photonics_2013}%
  \BibitemOpen
  \bibfield  {author} {\bibinfo {author} {\bibfnamefont {X.-D.}\ \bibnamefont
  {Cai}}, \bibinfo {author} {\bibfnamefont {C.}~\bibnamefont {Weedbrook}},
  \bibinfo {author} {\bibfnamefont {Z.-E.}\ \bibnamefont {Su}}, \bibinfo
  {author} {\bibfnamefont {M.-C.}\ \bibnamefont {Chen}}, \bibinfo {author}
  {\bibfnamefont {M.}~\bibnamefont {Gu}}, \bibinfo {author} {\bibfnamefont
  {M.-J.}\ \bibnamefont {Zhu}}, \bibinfo {author} {\bibfnamefont
  {L.}~\bibnamefont {Li}}, \bibinfo {author} {\bibfnamefont {N.-L.}\
  \bibnamefont {Liu}}, \bibinfo {author} {\bibfnamefont {C.-Y.}\ \bibnamefont
  {Lu}}, \ and\ \bibinfo {author} {\bibfnamefont {J.-W.}\ \bibnamefont {Pan}},\
  }\href {\doibase 10.1103/PhysRevLett.110.230501} {\bibfield  {journal}
  {\bibinfo  {journal} {Phys. Rev. Lett.}\ }\textbf {\bibinfo {volume} {110}},\
  \bibinfo {pages} {230501} (\bibinfo {year} {2013})}\BibitemShut {NoStop}%
\bibitem [{\citenamefont {Barz}\ \emph {et~al.}(2014)\citenamefont {Barz},
  \citenamefont {Kassal}, \citenamefont {Ringbauer}, \citenamefont {Lipp},
  \citenamefont {Daki{\ifmmode\acute{c}\else\'{c}\fi}}, \citenamefont
  {Aspuru-Guzik},\ and\ \citenamefont {Walther}}]{aspuru_photonics2_2014}%
  \BibitemOpen
  \bibfield  {author} {\bibinfo {author} {\bibfnamefont {S.}~\bibnamefont
  {Barz}}, \bibinfo {author} {\bibfnamefont {I.}~\bibnamefont {Kassal}},
  \bibinfo {author} {\bibfnamefont {M.}~\bibnamefont {Ringbauer}}, \bibinfo
  {author} {\bibfnamefont {Y.~O.}\ \bibnamefont {Lipp}}, \bibinfo {author}
  {\bibfnamefont {B.}~\bibnamefont {Daki{\ifmmode\acute{c}\else\'{c}\fi}}},
  \bibinfo {author} {\bibfnamefont {A.}~\bibnamefont {Aspuru-Guzik}}, \ and\
  \bibinfo {author} {\bibfnamefont {P.}~\bibnamefont {Walther}},\ }\href
  {\doibase 10.1038/srep06115} {\bibfield  {journal} {\bibinfo  {journal} {Sci.
  Rep.}\ }\textbf {\bibinfo {volume} {4}},\ \bibinfo {pages} {1} (\bibinfo
  {year} {2014})}\BibitemShut {NoStop}%
\bibitem [{\citenamefont {Duan}\ \emph {et~al.}(2020)\citenamefont {Duan},
  \citenamefont {Yuan}, \citenamefont {Yu}, \citenamefont {Huang},\ and\
  \citenamefont {Hsieh}}]{Supercond_hhl_2020}%
  \BibitemOpen
  \bibfield  {author} {\bibinfo {author} {\bibfnamefont {B.}~\bibnamefont
  {Duan}}, \bibinfo {author} {\bibfnamefont {J.}~\bibnamefont {Yuan}}, \bibinfo
  {author} {\bibfnamefont {C.-H.}\ \bibnamefont {Yu}}, \bibinfo {author}
  {\bibfnamefont {J.}~\bibnamefont {Huang}}, \ and\ \bibinfo {author}
  {\bibfnamefont {C.-Y.}\ \bibnamefont {Hsieh}},\ }\href {\doibase
  10.1016/j.physleta.2020.126595} {\bibfield  {journal} {\bibinfo  {journal}
  {Phys. Lett. A}\ }\textbf {\bibinfo {volume} {384}},\ \bibinfo {pages}
  {126595} (\bibinfo {year} {2020})}\BibitemShut {NoStop}%
\bibitem [{\citenamefont {Sen(De)}\ and\ \citenamefont
  {Sen}(2010)}]{sen_pra_2010}%
  \BibitemOpen
  \bibfield  {author} {\bibinfo {author} {\bibfnamefont {A.}~\bibnamefont
  {Sen(De)}}\ and\ \bibinfo {author} {\bibfnamefont {U.}~\bibnamefont {Sen}},\
  }\href {\doibase 10.1103/PhysRevA.81.012308} {\bibfield  {journal} {\bibinfo
  {journal} {Phys. Rev. A}\ }\textbf {\bibinfo {volume} {81}},\ \bibinfo
  {pages} {012308} (\bibinfo {year} {2010})}\BibitemShut {NoStop}%
\bibitem [{\citenamefont {Maiti}\ \emph {et~al.}(2022)\citenamefont {Maiti},
  \citenamefont {Sen},\ and\ \citenamefont {Sen}}]{maiti_arxiv_2022}%
  \BibitemOpen
  \bibfield  {author} {\bibinfo {author} {\bibfnamefont {S.}~\bibnamefont
  {Maiti}}, \bibinfo {author} {\bibfnamefont {K.}~\bibnamefont {Sen}}, \ and\
  \bibinfo {author} {\bibfnamefont {U.}~\bibnamefont {Sen}},\ }\href@noop {}
  {\enquote {\bibinfo {title} {Quantum phase estimation in presence of glassy
  disorder},}\ } (\bibinfo {year} {2022}),\ \Eprint
  {http://arxiv.org/abs/2112.04411} {arXiv:2112.04411 [quant-ph]} \BibitemShut
  {NoStop}%
\bibitem [{\citenamefont {Gupta}\ \emph {et~al.}(2023)\citenamefont {Gupta},
  \citenamefont {Ghosh}, \citenamefont {Sen},\ and\ \citenamefont
  {Sen}}]{gupta_arxiv_2023}%
  \BibitemOpen
  \bibfield  {author} {\bibinfo {author} {\bibfnamefont {A.}~\bibnamefont
  {Gupta}}, \bibinfo {author} {\bibfnamefont {P.}~\bibnamefont {Ghosh}},
  \bibinfo {author} {\bibfnamefont {K.}~\bibnamefont {Sen}}, \ and\ \bibinfo
  {author} {\bibfnamefont {U.}~\bibnamefont {Sen}},\ }\href@noop {} {\enquote
  {\bibinfo {title} {Effects of noise on performance of bernstein-vazirani
  algorithm},}\ } (\bibinfo {year} {2023}),\ \Eprint
  {http://arxiv.org/abs/2305.19745} {arXiv:2305.19745 [quant-ph]} \BibitemShut
  {NoStop}%
\bibitem [{\citenamefont {Berry}\ \emph {et~al.}(2007)\citenamefont {Berry},
  \citenamefont {Ahokas}, \citenamefont {Cleve},\ and\ \citenamefont
  {Sanders}}]{Berry_cmp_2007}%
  \BibitemOpen
  \bibfield  {author} {\bibinfo {author} {\bibfnamefont {D.~W.}\ \bibnamefont
  {Berry}}, \bibinfo {author} {\bibfnamefont {G.}~\bibnamefont {Ahokas}},
  \bibinfo {author} {\bibfnamefont {R.}~\bibnamefont {Cleve}}, \ and\ \bibinfo
  {author} {\bibfnamefont {B.~C.}\ \bibnamefont {Sanders}},\ }\href {\doibase
  10.1007/s00220-006-0150-x} {\bibfield  {journal} {\bibinfo  {journal}
  {Communications in Mathematical Physics}\ }\textbf {\bibinfo {volume}
  {270}},\ \bibinfo {pages} {359} (\bibinfo {year} {2007})}\BibitemShut
  {NoStop}%
\bibitem [{\citenamefont {Raeisi}\ \emph {et~al.}(2012)\citenamefont {Raeisi},
  \citenamefont {Wiebe},\ and\ \citenamefont {Sanders}}]{Barry_dynamics_2}%
  \BibitemOpen
  \bibfield  {author} {\bibinfo {author} {\bibfnamefont {S.}~\bibnamefont
  {Raeisi}}, \bibinfo {author} {\bibfnamefont {N.}~\bibnamefont {Wiebe}}, \
  and\ \bibinfo {author} {\bibfnamefont {B.~C.}\ \bibnamefont {Sanders}},\
  }\href {\doibase 10.1088/1367-2630/14/10/103017} {\bibfield  {journal}
  {\bibinfo  {journal} {New J. Phys.}\ }\textbf {\bibinfo {volume} {14}},\
  \bibinfo {pages} {103017} (\bibinfo {year} {2012})}\BibitemShut {NoStop}%
\bibitem [{\citenamefont {Wiebe}\ \emph {et~al.}(2011)\citenamefont {Wiebe},
  \citenamefont {Berry}, \citenamefont {H{\o}yer},\ and\ \citenamefont
  {Sanders}}]{Barry_dynamics_3}%
  \BibitemOpen
  \bibfield  {author} {\bibinfo {author} {\bibfnamefont {N.}~\bibnamefont
  {Wiebe}}, \bibinfo {author} {\bibfnamefont {D.~W.}\ \bibnamefont {Berry}},
  \bibinfo {author} {\bibfnamefont {P.}~\bibnamefont {H{\o}yer}}, \ and\
  \bibinfo {author} {\bibfnamefont {B.~C.}\ \bibnamefont {Sanders}},\ }\href
  {\doibase 10.1088/1751-8113/44/44/445308} {\bibfield  {journal} {\bibinfo
  {journal} {J. Phys. A: Math. Theor.}\ }\textbf {\bibinfo {volume} {44}},\
  \bibinfo {pages} {445308} (\bibinfo {year} {2011})}\BibitemShut {NoStop}%
\bibitem [{\citenamefont {Babukhin}(2023)}]{babukhin_pra_2023}%
  \BibitemOpen
  \bibfield  {author} {\bibinfo {author} {\bibfnamefont {D.~V.}\ \bibnamefont
  {Babukhin}},\ }\href {\doibase 10.1103/PhysRevA.107.042408} {\bibfield
  {journal} {\bibinfo  {journal} {Phys. Rev. A}\ }\textbf {\bibinfo {volume}
  {107}},\ \bibinfo {pages} {042408} (\bibinfo {year} {2023})}\BibitemShut
  {NoStop}%
\bibitem [{\citenamefont {Barnum}\ and\ \citenamefont
  {Linden}(2001)}]{ggm_Barnum2001Aug}%
  \BibitemOpen
  \bibfield  {author} {\bibinfo {author} {\bibfnamefont {H.}~\bibnamefont
  {Barnum}}\ and\ \bibinfo {author} {\bibfnamefont {N.}~\bibnamefont
  {Linden}},\ }\href {\doibase 10.1088/0305-4470/34/35/305} {\bibfield
  {journal} {\bibinfo  {journal} {J. Phys. A: Math. Gen.}\ }\textbf {\bibinfo
  {volume} {34}},\ \bibinfo {pages} {6787} (\bibinfo {year}
  {2001})}\BibitemShut {NoStop}%
\bibitem [{\citenamefont {Wei}\ and\ \citenamefont
  {Goldbart}(2003)}]{ggm_Goldbart2003}%
  \BibitemOpen
  \bibfield  {author} {\bibinfo {author} {\bibfnamefont {T.-C.}\ \bibnamefont
  {Wei}}\ and\ \bibinfo {author} {\bibfnamefont {P.~M.}\ \bibnamefont
  {Goldbart}},\ }\href {\doibase 10.1103/PhysRevA.68.042307} {\bibfield
  {journal} {\bibinfo  {journal} {Phys. Rev. A}\ }\textbf {\bibinfo {volume}
  {68}},\ \bibinfo {pages} {042307} (\bibinfo {year} {2003})}\BibitemShut
  {NoStop}%
\bibitem [{\citenamefont {Shimony}(1995)}]{ggm_Shimony1995Apr}%
  \BibitemOpen
  \bibfield  {author} {\bibinfo {author} {\bibfnamefont {A.}~\bibnamefont
  {Shimony}},\ }\href {\doibase 10.1111/j.1749-6632.1995.tb39008.x} {\bibfield
  {journal} {\bibinfo  {journal} {Ann. N.Y. Acad. Sci.}\ }\textbf {\bibinfo
  {volume} {755}},\ \bibinfo {pages} {675} (\bibinfo {year}
  {1995})}\BibitemShut {NoStop}%
\bibitem [{\citenamefont {Vidal}\ and\ \citenamefont
  {Werner}(2002{\natexlab{a}})}]{vidal_logneg_2002}%
  \BibitemOpen
  \bibfield  {author} {\bibinfo {author} {\bibfnamefont {G.}~\bibnamefont
  {Vidal}}\ and\ \bibinfo {author} {\bibfnamefont {R.~F.}\ \bibnamefont
  {Werner}},\ }\href {\doibase 10.1103/PhysRevA.65.032314} {\bibfield
  {journal} {\bibinfo  {journal} {Phys. Rev. A}\ }\textbf {\bibinfo {volume}
  {65}},\ \bibinfo {pages} {032314} (\bibinfo {year}
  {2002}{\natexlab{a}})}\BibitemShut {NoStop}%
\bibitem [{\citenamefont {Plenio}(2005)}]{plenio2005}%
  \BibitemOpen
  \bibfield  {author} {\bibinfo {author} {\bibfnamefont {M.~B.}\ \bibnamefont
  {Plenio}},\ }\href {\doibase 10.1103/PhysRevLett.95.090503} {\bibfield
  {journal} {\bibinfo  {journal} {Phys. Rev. Lett.}\ }\textbf {\bibinfo
  {volume} {95}},\ \bibinfo {pages} {090503} (\bibinfo {year}
  {2005})}\BibitemShut {NoStop}%
\bibitem [{\citenamefont {Baumgratz}\ \emph {et~al.}(2014)\citenamefont
  {Baumgratz}, \citenamefont {Cramer},\ and\ \citenamefont
  {Plenio}}]{baumgratz_prl_2014}%
  \BibitemOpen
  \bibfield  {author} {\bibinfo {author} {\bibfnamefont {T.}~\bibnamefont
  {Baumgratz}}, \bibinfo {author} {\bibfnamefont {M.}~\bibnamefont {Cramer}}, \
  and\ \bibinfo {author} {\bibfnamefont {M.~B.}\ \bibnamefont {Plenio}},\
  }\href {\doibase 10.1103/PhysRevLett.113.140401} {\bibfield  {journal}
  {\bibinfo  {journal} {Phys. Rev. Lett.}\ }\textbf {\bibinfo {volume} {113}},\
  \bibinfo {pages} {140401} (\bibinfo {year} {2014})}\BibitemShut {NoStop}%
\bibitem [{\citenamefont {Horodecki}\ \emph {et~al.}(1996)\citenamefont
  {Horodecki}, \citenamefont {Horodecki},\ and\ \citenamefont
  {Horodecki}}]{horodecki_pla_1996}%
  \BibitemOpen
  \bibfield  {author} {\bibinfo {author} {\bibfnamefont {M.}~\bibnamefont
  {Horodecki}}, \bibinfo {author} {\bibfnamefont {P.}~\bibnamefont
  {Horodecki}}, \ and\ \bibinfo {author} {\bibfnamefont {R.}~\bibnamefont
  {Horodecki}},\ }\href {\doibase
  https://doi.org/10.1016/S0375-9601(96)00706-2} {\bibfield  {journal}
  {\bibinfo  {journal} {Physics Letters A}\ }\textbf {\bibinfo {volume}
  {223}},\ \bibinfo {pages} {1} (\bibinfo {year} {1996})}\BibitemShut {NoStop}%
\bibitem [{\citenamefont {Peres}(1996)}]{peres_prl_1996}%
  \BibitemOpen
  \bibfield  {author} {\bibinfo {author} {\bibfnamefont {A.}~\bibnamefont
  {Peres}},\ }\href {\doibase 10.1103/PhysRevLett.77.1413} {\bibfield
  {journal} {\bibinfo  {journal} {Phys. Rev. Lett.}\ }\textbf {\bibinfo
  {volume} {77}},\ \bibinfo {pages} {1413} (\bibinfo {year}
  {1996})}\BibitemShut {NoStop}%
\bibitem [{\citenamefont {Vidal}\ and\ \citenamefont
  {Werner}(2002{\natexlab{b}})}]{vidal_pra_2002}%
  \BibitemOpen
  \bibfield  {author} {\bibinfo {author} {\bibfnamefont {G.}~\bibnamefont
  {Vidal}}\ and\ \bibinfo {author} {\bibfnamefont {R.~F.}\ \bibnamefont
  {Werner}},\ }\href {\doibase 10.1103/PhysRevA.65.032314} {\bibfield
  {journal} {\bibinfo  {journal} {Phys. Rev. A}\ }\textbf {\bibinfo {volume}
  {65}},\ \bibinfo {pages} {032314} (\bibinfo {year}
  {2002}{\natexlab{b}})}\BibitemShut {NoStop}%
\bibitem [{\citenamefont {Horodecki}\ \emph {et~al.}(1997)\citenamefont
  {Horodecki}, \citenamefont {Horodecki},\ and\ \citenamefont
  {Horodecki}}]{horodeckidistillable}%
  \BibitemOpen
  \bibfield  {author} {\bibinfo {author} {\bibfnamefont {M.}~\bibnamefont
  {Horodecki}}, \bibinfo {author} {\bibfnamefont {P.}~\bibnamefont
  {Horodecki}}, \ and\ \bibinfo {author} {\bibfnamefont {R.}~\bibnamefont
  {Horodecki}},\ }\href {\doibase 10.1103/PhysRevLett.78.574} {\bibfield
  {journal} {\bibinfo  {journal} {Phys. Rev. Lett.}\ }\textbf {\bibinfo
  {volume} {78}},\ \bibinfo {pages} {574} (\bibinfo {year} {1997})}\BibitemShut
  {NoStop}%
\bibitem [{\citenamefont {Horodecki}\ and\ \citenamefont
  {Horodecki}(1999)}]{horodecki_ent_distillable_1999}%
  \BibitemOpen
  \bibfield  {author} {\bibinfo {author} {\bibfnamefont {M.}~\bibnamefont
  {Horodecki}}\ and\ \bibinfo {author} {\bibfnamefont {P.}~\bibnamefont
  {Horodecki}},\ }\href {\doibase 10.1103/PhysRevA.59.4206} {\bibfield
  {journal} {\bibinfo  {journal} {Phys. Rev. A}\ }\textbf {\bibinfo {volume}
  {59}},\ \bibinfo {pages} {4206} (\bibinfo {year} {1999})}\BibitemShut
  {NoStop}%
\bibitem [{\citenamefont {Garnerone}\ \emph {et~al.}(2009)\citenamefont
  {Garnerone}, \citenamefont {Jacobson}, \citenamefont {Haas},\ and\
  \citenamefont {Zanardi}}]{garnerone_prl_2009}%
  \BibitemOpen
  \bibfield  {author} {\bibinfo {author} {\bibfnamefont {S.}~\bibnamefont
  {Garnerone}}, \bibinfo {author} {\bibfnamefont {N.~T.}\ \bibnamefont
  {Jacobson}}, \bibinfo {author} {\bibfnamefont {S.}~\bibnamefont {Haas}}, \
  and\ \bibinfo {author} {\bibfnamefont {P.}~\bibnamefont {Zanardi}},\ }\href
  {\doibase 10.1103/PhysRevLett.102.057205} {\bibfield  {journal} {\bibinfo
  {journal} {Phys. Rev. Lett.}\ }\textbf {\bibinfo {volume} {102}},\ \bibinfo
  {pages} {057205} (\bibinfo {year} {2009})}\BibitemShut {NoStop}%
\bibitem [{\citenamefont {Jacobson}\ \emph {et~al.}(2009)\citenamefont
  {Jacobson}, \citenamefont {Garnerone}, \citenamefont {Haas},\ and\
  \citenamefont {Zanardi}}]{jacobson_prb_2009}%
  \BibitemOpen
  \bibfield  {author} {\bibinfo {author} {\bibfnamefont {N.~T.}\ \bibnamefont
  {Jacobson}}, \bibinfo {author} {\bibfnamefont {S.}~\bibnamefont {Garnerone}},
  \bibinfo {author} {\bibfnamefont {S.}~\bibnamefont {Haas}}, \ and\ \bibinfo
  {author} {\bibfnamefont {P.}~\bibnamefont {Zanardi}},\ }\href {\doibase
  10.1103/PhysRevB.79.184427} {\bibfield  {journal} {\bibinfo  {journal} {Phys.
  Rev. B}\ }\textbf {\bibinfo {volume} {79}},\ \bibinfo {pages} {184427}
  (\bibinfo {year} {2009})}\BibitemShut {NoStop}%
\bibitem [{\citenamefont {Niederberger}\ \emph {et~al.}(2010)\citenamefont
  {Niederberger}, \citenamefont {Rams}, \citenamefont {Dziarmaga},
  \citenamefont {Cucchietti}, \citenamefont {Wehr},\ and\ \citenamefont
  {Lewenstein}}]{niederberger_pra_2010}%
  \BibitemOpen
  \bibfield  {author} {\bibinfo {author} {\bibfnamefont {A.}~\bibnamefont
  {Niederberger}}, \bibinfo {author} {\bibfnamefont {M.~M.}\ \bibnamefont
  {Rams}}, \bibinfo {author} {\bibfnamefont {J.}~\bibnamefont {Dziarmaga}},
  \bibinfo {author} {\bibfnamefont {F.~M.}\ \bibnamefont {Cucchietti}},
  \bibinfo {author} {\bibfnamefont {J.}~\bibnamefont {Wehr}}, \ and\ \bibinfo
  {author} {\bibfnamefont {M.}~\bibnamefont {Lewenstein}},\ }\href {\doibase
  10.1103/PhysRevA.82.013630} {\bibfield  {journal} {\bibinfo  {journal} {Phys.
  Rev. A}\ }\textbf {\bibinfo {volume} {82}},\ \bibinfo {pages} {013630}
  (\bibinfo {year} {2010})}\BibitemShut {NoStop}%
\bibitem [{\citenamefont {Sadhukhan}\ \emph {et~al.}(2016)\citenamefont
  {Sadhukhan}, \citenamefont {Prabhu}, \citenamefont {Sen(De)},\ and\
  \citenamefont {Sen}}]{sadhukhan_disorder_2016}%
  \BibitemOpen
  \bibfield  {author} {\bibinfo {author} {\bibfnamefont {D.}~\bibnamefont
  {Sadhukhan}}, \bibinfo {author} {\bibfnamefont {R.}~\bibnamefont {Prabhu}},
  \bibinfo {author} {\bibfnamefont {A.}~\bibnamefont {Sen(De)}}, \ and\
  \bibinfo {author} {\bibfnamefont {U.}~\bibnamefont {Sen}},\ }\href {\doibase
  10.1103/PhysRevE.93.032115} {\bibfield  {journal} {\bibinfo  {journal} {Phys.
  Rev. E}\ }\textbf {\bibinfo {volume} {93}},\ \bibinfo {pages} {032115}
  (\bibinfo {year} {2016})}\BibitemShut {NoStop}%
\bibitem [{\citenamefont {D'Alessio}\ \emph {et~al.}(2016)\citenamefont
  {D'Alessio}, \citenamefont {Kafri}, \citenamefont {Polkovnikov},\ and\
  \citenamefont {Rigol}}]{anatoli_adv_2016}%
  \BibitemOpen
  \bibfield  {author} {\bibinfo {author} {\bibfnamefont {L.}~\bibnamefont
  {D'Alessio}}, \bibinfo {author} {\bibfnamefont {Y.}~\bibnamefont {Kafri}},
  \bibinfo {author} {\bibfnamefont {A.}~\bibnamefont {Polkovnikov}}, \ and\
  \bibinfo {author} {\bibfnamefont {M.}~\bibnamefont {Rigol}},\ }\href
  {\doibase 10.1080/00018732.2016.1198134} {\bibfield  {journal} {\bibinfo
  {journal} {Advances in Physics}\ }\textbf {\bibinfo {volume} {65}},\ \bibinfo
  {pages} {239} (\bibinfo {year} {2016})},\ \Eprint
  {http://arxiv.org/abs/https://doi.org/10.1080/00018732.2016.1198134}
  {https://doi.org/10.1080/00018732.2016.1198134} \BibitemShut {NoStop}%
\bibitem [{\citenamefont {Konar}\ \emph {et~al.}(2022)\citenamefont {Konar},
  \citenamefont {Ghosh}, \citenamefont {Pal},\ and\ \citenamefont
  {Sen(De)}}]{konar_pra_2022}%
  \BibitemOpen
  \bibfield  {author} {\bibinfo {author} {\bibfnamefont {T.~K.}\ \bibnamefont
  {Konar}}, \bibinfo {author} {\bibfnamefont {S.}~\bibnamefont {Ghosh}},
  \bibinfo {author} {\bibfnamefont {A.~K.}\ \bibnamefont {Pal}}, \ and\
  \bibinfo {author} {\bibfnamefont {A.}~\bibnamefont {Sen(De)}},\ }\href
  {\doibase 10.1103/PhysRevA.105.022214} {\bibfield  {journal} {\bibinfo
  {journal} {Phys. Rev. A}\ }\textbf {\bibinfo {volume} {105}},\ \bibinfo
  {pages} {022214} (\bibinfo {year} {2022})}\BibitemShut {NoStop}%
\bibitem [{\citenamefont {Halder}\ \emph {et~al.}(2022)\citenamefont {Halder},
  \citenamefont {Banerjee}, \citenamefont {Ghosh}, \citenamefont {Pal},\ and\
  \citenamefont {Sen(De)}}]{halder_pra_2022}%
  \BibitemOpen
  \bibfield  {author} {\bibinfo {author} {\bibfnamefont {P.}~\bibnamefont
  {Halder}}, \bibinfo {author} {\bibfnamefont {R.}~\bibnamefont {Banerjee}},
  \bibinfo {author} {\bibfnamefont {S.}~\bibnamefont {Ghosh}}, \bibinfo
  {author} {\bibfnamefont {A.~K.}\ \bibnamefont {Pal}}, \ and\ \bibinfo
  {author} {\bibfnamefont {A.}~\bibnamefont {Sen(De)}},\ }\href {\doibase
  10.1103/PhysRevA.106.032604} {\bibfield  {journal} {\bibinfo  {journal}
  {Phys. Rev. A}\ }\textbf {\bibinfo {volume} {106}},\ \bibinfo {pages}
  {032604} (\bibinfo {year} {2022})}\BibitemShut {NoStop}%
\bibitem [{\citenamefont {Muhuri}\ \emph {et~al.}(2022)\citenamefont {Muhuri},
  \citenamefont {Gupta}, \citenamefont {Ghosh},\ and\ \citenamefont
  {De}}]{muhuri_arxiv_2022}%
  \BibitemOpen
  \bibfield  {author} {\bibinfo {author} {\bibfnamefont {A.}~\bibnamefont
  {Muhuri}}, \bibinfo {author} {\bibfnamefont {R.}~\bibnamefont {Gupta}},
  \bibinfo {author} {\bibfnamefont {S.}~\bibnamefont {Ghosh}}, \ and\ \bibinfo
  {author} {\bibfnamefont {A.~S.}\ \bibnamefont {De}},\ }\href@noop {}
  {\enquote {\bibinfo {title} {Superiority in dense coding through
  non-markovian stochasticity},}\ } (\bibinfo {year} {2022}),\ \Eprint
  {http://arxiv.org/abs/2211.13057} {arXiv:2211.13057 [quant-ph]} \BibitemShut
  {NoStop}%
\bibitem [{\citenamefont {Bhattacharyya}\ \emph {et~al.}(2023)\citenamefont
  {Bhattacharyya}, \citenamefont {Ghoshal},\ and\ \citenamefont
  {Sen}}]{bhattacharyya_arxiv_2023}%
  \BibitemOpen
  \bibfield  {author} {\bibinfo {author} {\bibfnamefont {A.}~\bibnamefont
  {Bhattacharyya}}, \bibinfo {author} {\bibfnamefont {A.}~\bibnamefont
  {Ghoshal}}, \ and\ \bibinfo {author} {\bibfnamefont {U.}~\bibnamefont
  {Sen}},\ }\href@noop {} {\enquote {\bibinfo {title} {Disorder-induced
  enhancement of precision in quantum metrology},}\ } (\bibinfo {year}
  {2023}),\ \Eprint {http://arxiv.org/abs/2212.08523} {arXiv:2212.08523
  [quant-ph]} \BibitemShut {NoStop}%
\bibitem [{\citenamefont {Monika}\ \emph {et~al.}(2023)\citenamefont {Monika},
  \citenamefont {Lakkaraju}, \citenamefont {Ghosh},\ and\ \citenamefont
  {De}}]{monika_arxiv_2023}%
  \BibitemOpen
  \bibfield  {author} {\bibinfo {author} {\bibnamefont {Monika}}, \bibinfo
  {author} {\bibfnamefont {L.~G.~C.}\ \bibnamefont {Lakkaraju}}, \bibinfo
  {author} {\bibfnamefont {S.}~\bibnamefont {Ghosh}}, \ and\ \bibinfo {author}
  {\bibfnamefont {A.~S.}\ \bibnamefont {De}},\ }\href@noop {} {\enquote
  {\bibinfo {title} {Better sensing with variable-range interactions},}\ }
  (\bibinfo {year} {2023}),\ \Eprint {http://arxiv.org/abs/2307.06901}
  {arXiv:2307.06901 [quant-ph]} \BibitemShut {NoStop}%
\end{thebibliography}%
\appendix


\section{Genuine multipartite entanglement in all bipartitions: Micro GGM }

Let us discuss an important aspect in the calculation of  GGM in the second step. We have considered the state made up of three subsystems: $\mathbf{\Lambda} : \mathbf{U} : \mathbf{R}$, where
the first subsystem encodes the eigenvalues, the second one encodes the eigenvectors and the third one is the auxiliary qubit. In the step 1 of the algorithm, the eigenvalues are encoded in the first subsystem ($\mathbf{\Lambda}$) as a string of qubit which represents the eigenvalues in bits, for example if the eigenvalue is \(5\),  the state $\ket{5}$ is actually $\ket{101} = \ket{1} \otimes \ket{0}  \otimes \ket{1}$. Now consider the same two-dimensional example as in Sec. \ref{sec:connect}, the states in the Step 2 with eigenvalues \(5(101)\) and \(7(111)\) is given  by
\begin{equation} \ket{\psi_{2}} = \sum_{i=1}^{N}\ket{\lambda_{i}} \otimes \beta_{i} \ket{u_{i}} \otimes (\sqrt{1-\frac{C^{2}}{\lambda_{i}^{2}}}\ket{0} + \frac{C}{\lambda_{i}}\ket{1}),
\end{equation}
which can be rewritten as
\begin{align*}
\ket{\psi_{2}} = \ket{\lambda_{1}} \otimes \beta_{1} \ket{u_{1}} \otimes \Big(\sqrt{1-\frac{C^{2}}{\lambda_{1}^{2}}}\ket{0} + \frac{C}{\lambda_{1}}\ket{1}\Big) \\
+ \ket{\lambda_{2}} \otimes \beta_{2} \ket{u_{2}} \otimes \Big(\sqrt{1-\frac{C^{2}}{\lambda_{2}^{2}}}\ket{0} + \frac{C}{\lambda_{2}}\ket{1}\Big).
\end{align*}
Substituting the eigenvalues by quantum states, the modified state reads as
\begin{align*}
\ket{\psi_{2}} = \ket{101} \otimes \beta_{1} \ket{u_{1}} \otimes \Big(\sqrt{1-\frac{C^{2}}{\lambda_{1}^{2}}}\ket{0} + \frac{C}{\lambda_{1}}\ket{1}\Big) \\
+ \ket{111} \otimes \beta_{2} \ket{u_{2}} \otimes \Big(\sqrt{1-\frac{C^{2}}{\lambda_{2}^{2}}}\ket{0} + \frac{C}{\lambda_{2}}\ket{1}\Big).
\end{align*}
Factoring out  $\ket{1}$, we can write it as as 
\begin{align*}
\ket{\psi_{2}} = \ket{1} \otimes \Big[ \ket{01} \otimes \beta_{1} \ket{u_{1}} \otimes \Big(\sqrt{1-\frac{C^{2}}{\lambda_{1}^{2}}}\ket{0} + \frac{C}{\lambda_{1}}\ket{1}\Big) \\
+ \ket{11} \otimes \beta_{2} \ket{u_{2}} \otimes \Big(\sqrt{1-\frac{C^{2}}{\lambda_{2}^{2}}}\ket{0} + \frac{C}{\lambda_{2}}\ket{1}\Big) \Big],
\end{align*}
having vanishing GME, i.e., \(\mathcal{E} (\ket{\psi_2}) =0\) if we go beyond the partition \(\mathbf{\Lambda: U:R}\). 
Hence we can say that although the micro-level GGM is zero, i.e., the entire \(N\)-party system does not have genuine multipartite entanglement, nonvanishing GGM of the tripartite system which is the building block for the algorithm only  plays a role.

\end{document}